\documentclass[times, review, 10pt]{elsarticle}

\usepackage{amssymb}
\usepackage{amsmath}
\usepackage{booktabs} 
\usepackage{url,multirow,subcaption}

\journal{Pattern Recognition}

\begin{document}

\begin{frontmatter}



\title{Benchmarking Multimodal Large Language Models for Missing Modality Completion in Product Catalogues} 

\author[1]{Junchen Fu}
\ead{j.fu.3@research.gla.ac.uk}

\author[1]{Wenhao Deng}
\ead{w.deng.1@research.gla.ac.uk}

\author[1]{Kaiwen Zheng}
\ead{k.zheng.1@research.gla.ac.uk}


\author[3]{Ioannis Arapakis}
\ead{arapakis.ioannis@gmail.com}

\author[1]{Yu Ye}
\ead{2993533y@student.gla.ac.uk}

\author[4]{Yongxin Ni}
\ead{niyongxin@u.nus.edu}

\author[1]{Joemon M. Jose}
\ead{Joemon.Jose@glasgow.ac.uk}

\author[5]{Xuri Ge\corref{cor1}}
\ead{xuri.ge@sdu.edu.cn}

\cortext[cor1]{Corresponding author.}

\affiliation[1]{organization={University of Glasgow},
            city={Glasgow},
            country={United Kingdom}}


\affiliation[3]{organization={Telefónica Scientific Research},
            city={Barcelona},
            country={Spain}}

\affiliation[4]{organization={National University of Singapore},
            city={Singapore},
            country={Singapore}}

\affiliation[5]{organization={Shandong University},
            city={Jinan},
            country={China}}

\begin{abstract}
Missing-modality information on e-commerce platforms—such as absent product images or textual descriptions—often arises from annotation errors or incomplete metadata, impairing both the product presentation phase and downstream applications like recommendation systems. Inspired by the multimodal generative abilities of recent Multimodal Large Language Models (MLLMs), this work investigates a fundamental and underexplored question: \textit{Can MLLMs  generate missing modalities for products in e-commerce scenarios?} We propose the \textbf{Missing Modality Product Completion Benchmark (MMPCBench)}, which comprises two sub-benchmarks: the \emph{Content Quality Completion Benchmark} and the \emph{Recommendation Benchmark}. 
In addition, we evaluated six state-of-the-art MLLMs from the Qwen2.5-VL and Gemma-3 families in nine real-world e-commerce categories, focusing on \emph{Image-to-Text} and \emph{Text-to-Image} completion tasks. 
Experiments show that MLLMs capture high-level semantics but struggle with fine-grained word- and pixel/patch-level alignment. 
Beyond this, we also find that performance varies by category and model scale, and there is no trivial correlation with model size as seen in mainstream benchmarks. We further explore \emph{Group Relative Policy Optimization} (GRPO) for a better alignment of the MLLMs for this task: it improves the Image-to-Text completion but does not improve the Text-to-Image completion. 
These findings highlight the limitations of current MLLMs in real-world cross-modal generation and mark an early step toward more effective missing-modality product completion.
\end{abstract}
\begin{keyword}
Missing Modality \sep MMPCBench \sep MLLMs \sep Product Catalogues
\end{keyword}

\end{frontmatter}



\section{Introduction}

\begin{figure}
    \centering
    \includegraphics[width=0.8\linewidth]{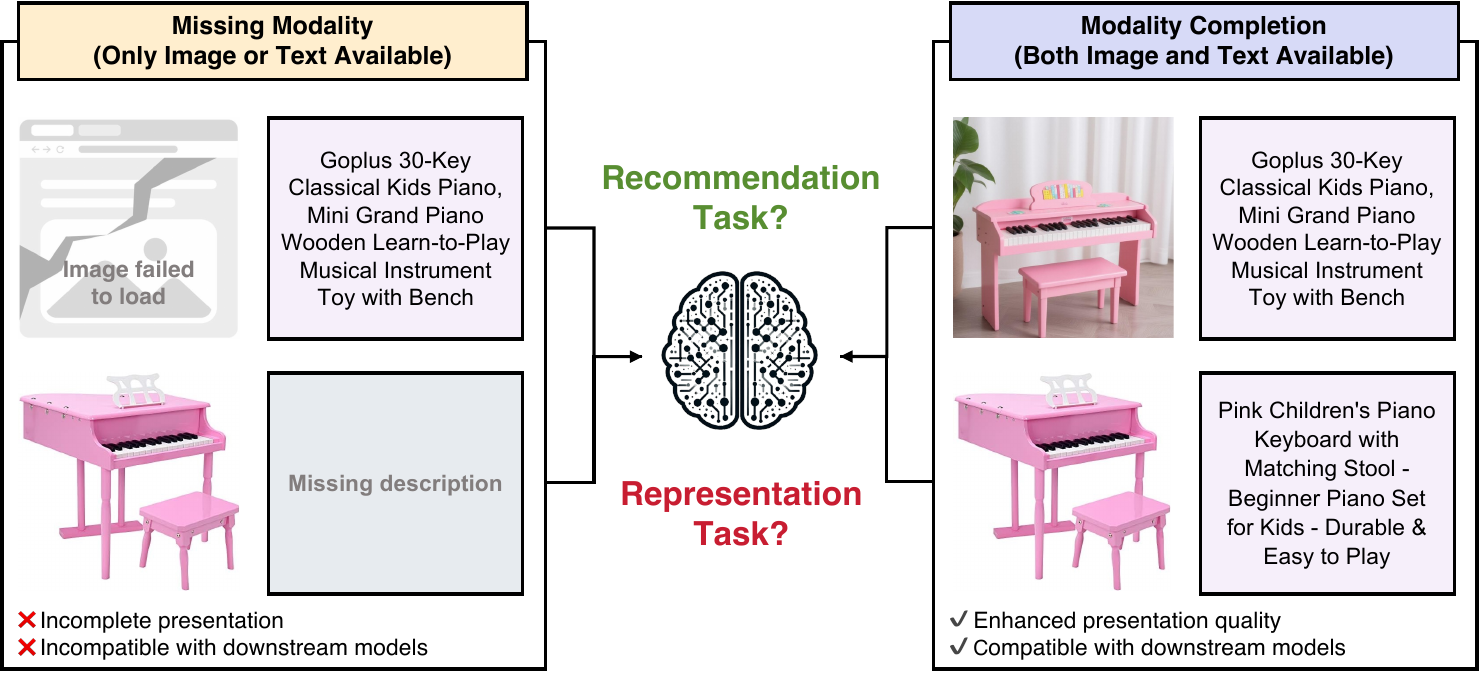}
    \caption{Missing-Modality Completion for E-Commerce Products: Given only one modality (either image or text), the task is to generate the missing counterpart. Missing modalities lead to incomplete product presentation, which can negatively impact downstream tasks such as representation learning and recommendation.}
    \label{fig:modality_completion}
 \vspace{-0.1in}
\end{figure}

Online marketplaces thrive on rich multimodal product content—photos that convey appearance and text that clarifies specifications, usage, and brand story \citep{li2020aspect}. In practice, however, catalog entries are frequently incomplete~\citep{malitesta2024we}: sellers omit images, provide terse or noisy descriptions, or mislabel key attributes, leaving significant portions of the inventory partially observed.\footnote{Although platforms have long mandated that sellers provide both product descriptions and images in compliance with strict regulations, some listings on major platforms such as Amazon still face these issues~\citep{zhu2021knowledge}. Possible causes include outdated standards, inconsistent practices among third-party sellers, broken images, corrupted text, or other technical factors. As a result, this remains an active area of research, particularly in the field of recommendation~\citep{wang2018lrmm,malitesta2024we}.} Such gaps propagate downstream, weakening retrieval, matching, and recommendation pipelines that depend on both visual and textual signals to infer user intent and item relevance. As recent work in multimodal recommendation and attribute extraction notes, missing or unreliable modalities are widespread and materially degrade system quality if left unaddressed \citep{wang2023mpkgac,fu2025crossan}. Therefore, completing the modality information provides more semantically meaningful user presentation and enables greater flexibility for downstream tasks, as illustrated in Figure~\ref{fig:modality_completion}.

Classical remedies—manual curation or statistical imputation (e.g., MICE \citep{vanbuuren2011mice}, MissForest \citep{stekhoven2012missforest})—are effective for low-dimensional tabular fields but fall short when the goal is to reconstruct semantically faithful images or free-form textual descriptions. Most existing approaches to handling missing modalities, especially in domains like e-commerce, operate primarily at the feature level—focusing on imputing embeddings or attributes rather than generating realistic content. Recently, however, the rise of Multimodal Large Language Models (MLLMs) like ChatGPT-4o, Qwen2.5‑VL, and Gemma‑3—has opened up promising new possibilities. In fields like medical imaging and scientific visualization, these models have already demonstrated impressive performance in reconstructing missing modalities from available inputs~\citep{ke2025knowledge}. Yet in the context of e-commerce, the dominant paradigm remains feature imputation, with limited exploration of MLLM-based generation.

To this end, we present the \textbf{Missing Modality Product Completion Benchmark (MMPCBench)}, comprising the \emph{Content Quality Completion Benchmark (CQBench)} and the \emph{Recommendation Benchmark (RecBench)}. Six state-of-the-art MLLMs from the Qwen2.5-VL and Gemma-3 families are evaluated on \emph{Image-to-Text} (I$\rightarrow$T) and \emph{Text-to-Image} (T$\rightarrow$I) completion tasks. Experiments reveal that performance varies across categories and model scales in a non-monotonic manner, exhibits \textit{task-wise} asymmetries between I$\rightarrow$T and T$\rightarrow$I, and shows only a weak correlation between content quality and recommendation performance.  To address misalignment of the existing MLLMs with e-commerce product completion, we explore \textbf{Group Relative Policy Optimization (GRPO)}, which improves I$\rightarrow$T but remains limited on T$\rightarrow$I. 

\textbf{Our contributions are:}
\begin{itemize}
    \item We introduce \textbf{MMPCBench}, the first benchmark for missing-modality product completion in e-commerce, comprising two sub-benchmarks: the \emph{CQBench} and the \emph{RecBench}, covering both I$\rightarrow$T and T$\rightarrow$I tasks across diverse product categories.
    \item We conduct a systematic evaluation of six state-of-the-art MLLMs, spanning three sizes each from the Qwen2.5‑VL and Gemma‑3 families, under a unified protocol for cross-modal completion.
    \item We explore \textbf{Group Relative Policy Optimization (GRPO)} for multimodal completion to enhance alignment of MLLMs in missing-modality completion.
\end{itemize}

\vspace{-0.1in}
\section{Related Work}
\paragraph{Missing Modality Completion in Multimodal Learning}
Multimodal learning often encounters scenarios where one or more data modalities are unavailable due to sensor failures, cost constraints, or data collection issues~\citep{ma2021smil}. This challenge has led to increasing interest in missing modality completion, which aims to enhance model robustness in the presence of incomplete modality inputs
. 
Early efforts approached this as a data imputation problem, relying on manual curation or classical statistical techniques such as MICE~\citep{vanbuuren2011mice} and MissForest~\citep{stekhoven2012missforest}. Neural network-based methods, including generative approaches like multimodal autoencoders and GANs, have also been proposed~\citep{ma2021smil}. Another line of studies moved away from explicit imputation, instead designing architectures that are inherently resilient to modality dropouts~\citep{ma2022robust}. For example, \citet{lee2023prompting} proposed multimodal prompting to compensate for missing modalities, while~\citet{wang2020transmodality} introduced a fusion mechanism to complete missing modalities for sentiment analysis tasks. \citet{chen2025adaptive} addresses the missing modality problem in multimodal Alzheimer's disease biomarker detection. MCE~\citep{zhao2025mce} handles missing modalities under imbalanced missing rates. MIMAR-OSA~\citep{qiu2025mimar} addresses obstructive sleep apnea diagnosis using multimodal data with missing modality reconstruction. However, these methods typically require training on complete multimodal data. Recent advances leverage large multimodal language models (MLLMs) to perform zero-shot missing modality completion in tasks such as visual question answering and healthcare~\citep{ke2025knowledge,fu2026llmpopcorn}, achieving state-of-the-art performance.  These encouraging results motivate us to explore the application of MLLMs for missing modality completion in the product domain.

\paragraph{Product Missing Modality Completion}
In e-commerce and product-centric applications~\cite{fu2024exploring}, a persistent issue is the incompleteness of multimodal product data—such as missing images, incomplete product descriptions, or absent attribute tags. This can result from cataloging errors, crowdsourcing inconsistencies, or metadata loss during system migrations. \citet{wang2018lrmm} pioneered latent regression-based methods to estimate missing product content, followed by adaptive feature sampling~\citep{shi2019adaptive} for handling incomplete product features. More recently, \citet{liu2022attribute} and \citet{malitesta2024we} propose graph-based feature propagation approaches to recover missing attributes or modality features for products in large-scale catalogs. These approaches leverage product–product similarity or co-purchase relations to impute realistic content. \citet{malitesta2024we} shows that dropping products with missing modalities is wasteful, and proposes modular imputation pipelines applicable before model training. \citet{kim2025dgmrec} advance this line by developing DGMRec, a generative model that synthesizes missing modalities using disentangled latent representations of available product features, enabling cold-start and zero-modality product recovery. However, mainstream existing approaches are all designed specifically for recommendation downstream tasks~\cite{fu2024iisan,fu2025efficient}, requiring task-specific training or fine-tuning, which limits their applicability to missing modality completion in other scenarios, such as presentation or alternative tasks.

Despite the growing interest in product modality completion and the promising potential of using Multimodal Large Language Models (MLLMs) for this task, no prior work has systematically explored leveraging advanced MLLMs—such as Qwen-VL and Gemma-3—for completing missing modalities in product data. MLLMs are inherently capable of integrating textual and visual contexts and generating missing components. However, their potential for product-level modality generation, especially in large-scale, real-world scenarios such as e-commerce catalogs, remains largely underexplored. This gap motivates our study, which presents the first comprehensive benchmark for the product missing modality completion task.

\begin{figure*}
    \centering
    \includegraphics[width=\linewidth]{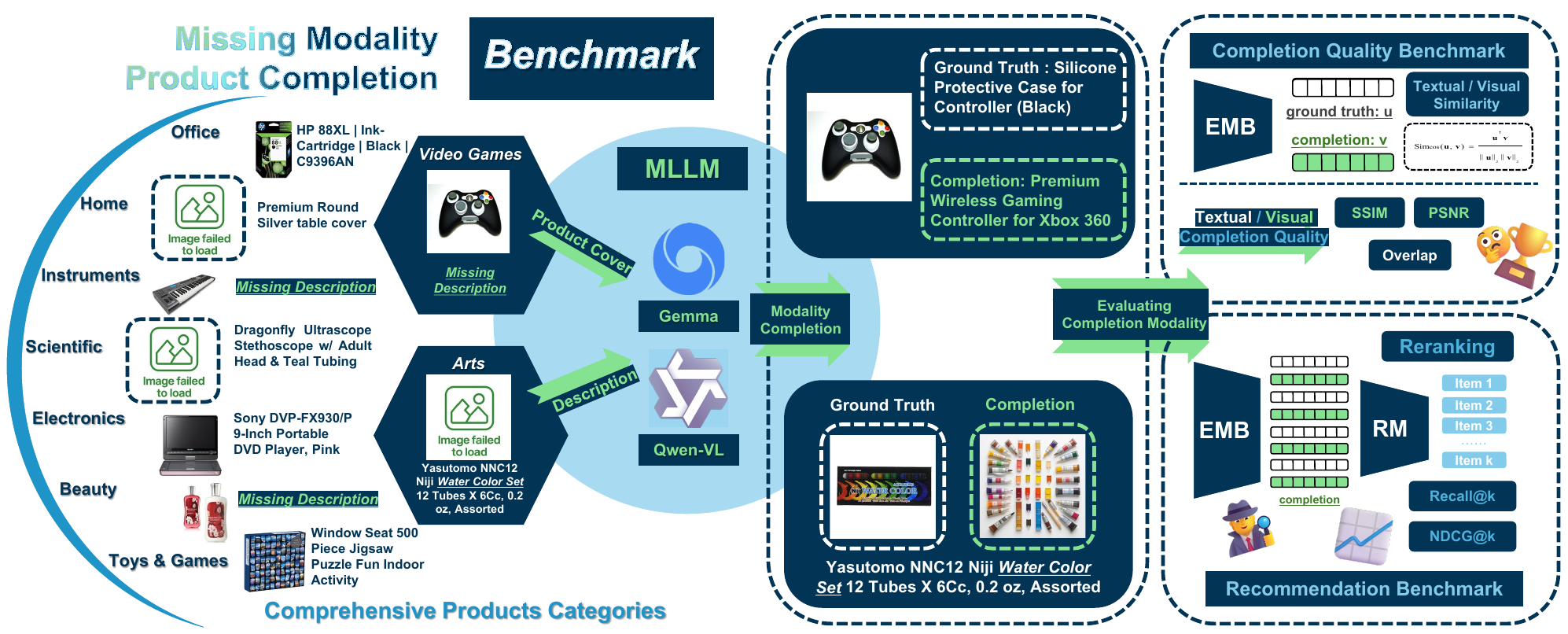}
    \caption{Overview of MMPCBench. MMPCBench is a benchmark designed to evaluate MLLMs on the task of completing missing product modalities—either text or image—across diverse product categories. EMB denotes the embedding models used for similarity-based evaluation, while RM refers to the recommender models employed for downstream recommendation tasks. The generated modalities are evaluated using these two components of the benchmark. Notably, completing missing images requires a diffusion model to visualize the image prompts generated by the MLLMs.}
    \label{fig:MMPCBench}
    \vspace{-0.1in}
\end{figure*}

\section{Methodology}

\subsection{Problem Formulation}
\label{sec:problem_formulation}
To identify the most critical elements for presentation and downstream tasks, we delimit our study to focus on the text description and product cover image, as they constitute the most essential components for on-site demonstration and downstream applications such as multimodal recommendation~\citep{zhou2023mmrec}.\footnote{Other attributes such as brand, price, and category are excluded, as their real-world values are more relevant than their visual or textual presentation. Due to the potential hallucinations of MLLMs when generating such attributes, we leave them for future exploration.}

We formulate the missing modality completion task using Multimodal Large Language Models (MLLMs) in an e-commerce context, where each product is represented as a multimodal pair \((x^T, x^I)\), consisting of a textual description \(x^T\) and a product image \(x^I\). However, in real-world scenarios, one modality may be missing due to annotation gaps or system constraints~\citep{malitesta2024we}. To address this, we define two complementary tasks: \textbf{Text-to-Image Completion}, which aims to generate an image \(\hat{x}^I\) from the available text \(x^T\). In this task, we leverage the product description as input to the MLLM to produce an image prompt, which is then passed to a Text-to-Image diffusion model to synthesize the image; and \textbf{Image-to-Text Completion}, which aims to generate a textual description \(\hat{x}^T\) from the product image \(x^I\), using the MLLM to translate visual information into coherent text.

To evaluate the effectiveness of modality completion, we adopt a dual-perspective framework. First, we assess \textit{presentation-level fidelity}, which measures the semantic alignment and perceptual quality of the generated modality. Second, we examine the \textit{recommendation-level impact}, which evaluates whether the generated modality can serve as an effective substitute for the missing input in downstream tasks such as personalized ranking. Specifically, we compare the recommendation performance using the generated modality against that of the original modality. This twofold evaluation ensures both the intrinsic plausibility and the extrinsic utility of the generated content. The detailed evaluation procedure is described in the following section.
\begin{table}[ht]
\centering
\small
\caption{Recommendation Benchmark Category Description. For the Content Quality Benchmark, we sampled 1,000 items per category.}
\begin{tabular}{lccc}
\toprule
Category Dataset & User & Item & Interactions  \\
\midrule
All Beauty & 1,068 &4,094 & 10,198  \\
Arts, Crafts \& Sewing &  8,692 & 4,250 & 55,714 \\
Electronics &  7,191 & 3,056 & 91,087 \\
Home \& Kitchen &  6,896 & 3,163 & 68,085 \\
Industrial \& Scientific &  5,437 & 4,034 & 33,026 \\
Musical Instruments &  12,890 & 3,640 & 91,603 \\
Office Products &  14,167 & 4,275 & 90,011 \\
Toys \& Games &  12,919 & 4,542 & 81,937 \\
Video Games &  25,501 & 3,461 & 189,692 \\
\bottomrule
\end{tabular}
\vspace{-0.1in}
\label{tab:recommendation_dataset}
\end{table}

\subsection{MMPCBench}
\label{sec:mmpcbench}
To systematically evaluate the capabilities of MLLMs in completing missing modalities for product data, we introduce MMPCBench, a benchmark specifically designed for missing modality product completion, as illustrated in Figure~\ref{fig:MMPCBench}. MMPCBench covers a comprehensive range of product categories to enable detailed and thorough evaluation. We adopt three different scales of open-sourced MLLMs from the state-of-the-art Gemma-3 and Qwen2.5-VL families. As defined in problem formulation section, our evaluation is twofold: the completion performance is assessed using two benchmarks, namely the \textit{Completion Quality Benchmark} and the \textit{Recommendation Benchmark}.

\paragraph{Dataset} To ensure a comprehensive evaluation, we adopt the largest e-commerce product dataset available on the Amazon platform. To mitigate potential data leakage---since many MLLMs may have been pre-trained on earlier versions of the Amazon Review dataset---we use the most recent version, the Amazon Review Dataset\footnote{\url{https://amazon-reviews-2023.github.io/}}, released in March 2024. Following prior works~\citep{hou2022towards, zhang2025hierarchical}, we focus on nine widely used product categories for multimodal recommendation: \textit{``All Beauty''}, \textit{``Arts, Crafts \& Sewing''}, \textit{``Electronics''}, \textit{``Home \& Kitchen''}, \textit{``Industrial \& Scientific''}, \textit{``Musical Instruments''}, \textit{``Office Products''}, \textit{``Toys \& Games''}, and \textit{``Video Games''}. As detailed in Table~\ref{tab:recommendation_dataset}, we follow the five-core filtering procedure from~\citep{zhou2023mmrec}, retaining only items with both textual description and product cover. For presentation benchmarking, we further sample 1,000 items from each category.

\paragraph{Evaluation}
MMPCBench is organized into two evaluation subcategories: \textit{Content Quality Benchmark} and \textit{Recommendation Benchmark}, which are designed to assess distinct aspects of modality completion performance in e-commerce scenarios. Each subcategory encompasses both \textit{Image-to-Text} and \textit{Text-to-Image} tasks.

\noindent In the \textit{Completion Quality Benchmark} (CQBench), for each sample, we mask either the product image or the textual description and use a Multimodal Large Language Model (MLLM), such as Qwen2.5-VL or Gemma-3, to complete the missing modality. For Text-to-Image tasks, the MLLMs are required to generate a product cover prompt, which is then fed into a diffusion-based model to produce the final images following~\citep{ke2025knowledge}. The quality of the generated modality is evaluated using similarity measures~\citep{shen2020multi} and standard automatic metrics~\citep{lan2025survey}. For each reported value, we present the mean along with its standard deviation.

For text completion, we evaluate: 
\textbf{Cosine Similarity}~\citep{tata2007estimating} – cosine of the angle between TF-IDF vectors of original and generated texts; 
\textbf{Euclidean Similarity}~\citep{elmore2001euclidean} — L2 distance on TF–IDF; add 1 and take the reciprocal for similarity; \textbf{Token Overlap}~\citep{wang2020measurement} – proportion of shared tokens, reflecting lexical similarity; 
\textbf{BERTScore}~\citep{zhang2019bertscore} – semantic similarity based on contextual BERT embeddings.

For image completion, we assess: 
\textbf{PSNR}~\citep{hore2010image} – pixel-level reconstruction quality; 
\textbf{SSIM}~\citep{wang2004image} – perceived image quality via luminance, contrast, and structure; 
\textbf{MSE}~\citep{tan2013perceptually} – mean squared pixel-wise difference; 
\textbf{LPIPS}~\citep{zhang2018unreasonable} – perceptual similarity using deep features; 
\textbf{CLIP Similarity}~\citep{radford2021clip} – cosine similarity of CLIP embeddings for semantic alignment.

\noindent In the \textit{Recommendation Benchmark}, we evaluate the effectiveness of the generated modality as a substitute in multimodal recommendation tasks. Specifically, we test whether replacing the original modality with the generated one affects recommendation performance, using three classic and popular multimodal recommendation models:

\begin{itemize}
    \item \textbf{VBPR} \citep{he2016vbpr}: A shallow model that extends matrix factorization by incorporating multimodal features extracted from product images and textual descriptions.
    
    \item \textbf{BM3} \citep{zhou2023bootstrap}: A self-supervised learning framework designed for multimodal recommendation.
    
    \item \textbf{FREEDOM} \citep{zhou2023tale}: A state-of-the-art multimodal recommender that simultaneously freezes the item-item graph and denoises the user-item interaction graph to enhance recommendation performance.
    
\end{itemize}

\noindent
We evaluate the performance difference of using the completed modality with top-$k$ metrics: 
\textbf{Recall@$k$} – proportion of relevant items in the top-$k$ list, and 
\textbf{NDCG@$k$} – relevance-aware metric giving higher scores to relevant items ranked earlier. We report results for $k = \{5, 10, 20, 50\}$ to evaluate performance across different recommendation list lengths. 

 Overall, MMPCBench provides a unified and practical testbed to evaluate both the fidelity of modality completion and its downstream impact on recommendation quality in realistic e-commerce scenarios. Implementation details of this paper are provided in the Appendix.

\begin{figure*}
    \centering
    \includegraphics[width=0.95\linewidth]{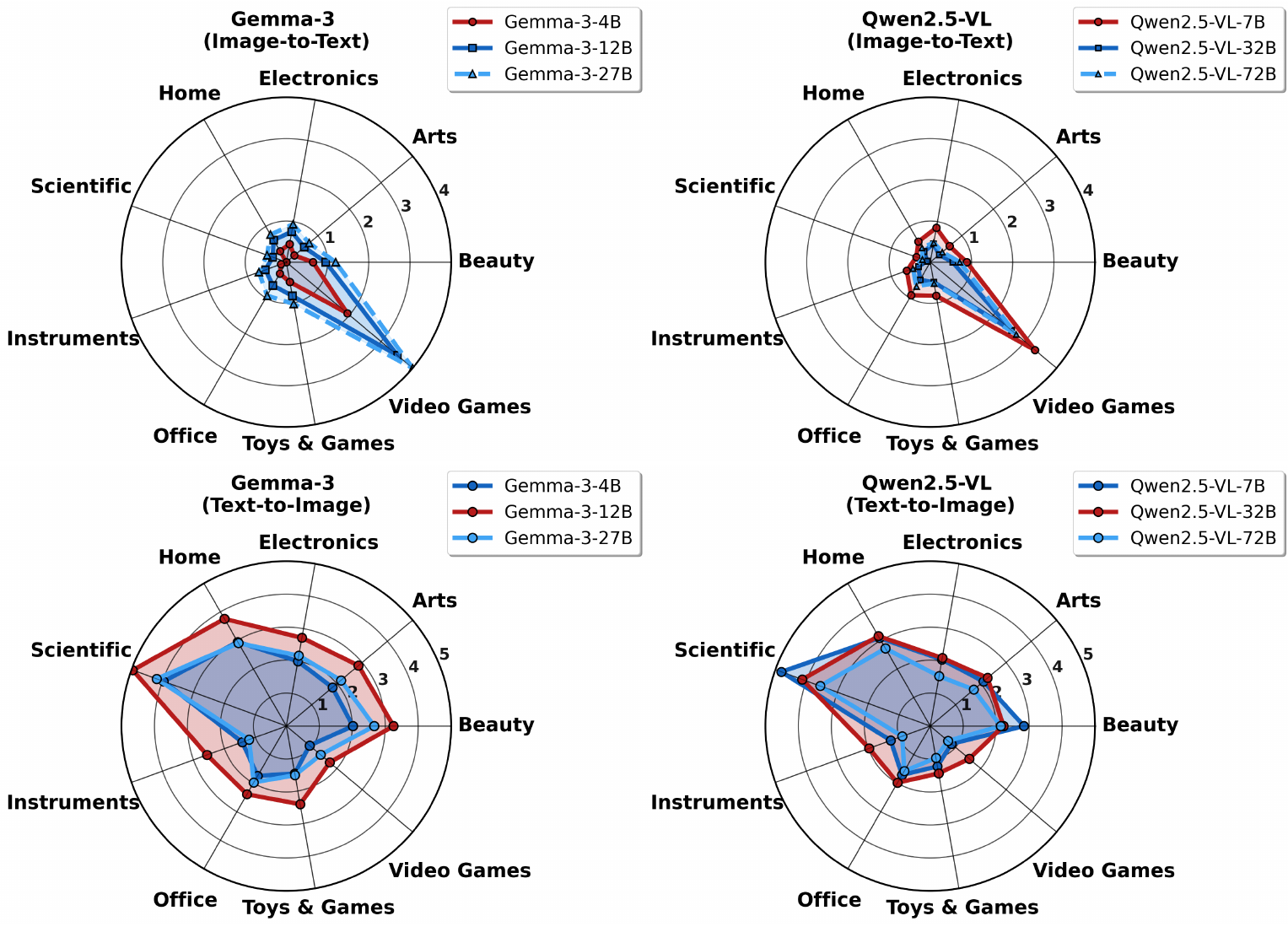}
    \caption{Radar Charts for Both T$\rightarrow$I and I$\rightarrow$T Tasks on Content Quality Benchmark. Each metric is normalized to a [0, 1] range and then summed, resulting in a maximum score of 4 for I$\rightarrow$T and 5 for T$\rightarrow$I tasks.}
    \vspace{-0.1in}
    \label{fig:radar_presentatoin}
    \vspace{-0.2in}
\end{figure*}

\begin{table*}[!ht]
\centering
\small
\setlength{\tabcolsep}{5pt}
\renewcommand{\arraystretch}{0.8}
\caption{Average Text Quality Metrics (mean $\pm$ std) Across All Categories for I$\rightarrow$T tasks on Content Quality Benchmark. $\uparrow$: larger is better. }
\begin{tabular}{lcccc}
\toprule
Model & Cosine $\uparrow$ & Euclidean $\uparrow$ & Overlap $\uparrow$ & BERTScore $\uparrow$ \\
\midrule
Gemma-3-4B & 0.166 $\pm$ 0.043 & 0.438 $\pm$ 0.008 & 0.213 $\pm$ 0.038 & 0.862 $\pm$ 0.008 \\
Gemma-3-12B & \underline{0.196 $\pm$ 0.062} & \underline{0.447 $\pm$ 0.022} & 0.251 $\pm$ 0.063 & 0.869 $\pm$ 0.012 \\
Gemma-3-27B & \textbf{0.210 $\pm$ 0.067} & \textbf{0.451 $\pm$ 0.028} & \textbf{0.269 $\pm$ 0.064} & \textbf{0.872 $\pm$ 0.012} \\
Qwen2.5-VL-7B-Instruct & 0.196 $\pm$ 0.061 & 0.446 $\pm$ 0.017 & \underline{0.254 $\pm$ 0.060} & \underline{0.870 $\pm$ 0.011} \\
Qwen2.5-VL-32B-Instruct & 0.169 $\pm$ 0.053 & 0.439 $\pm$ 0.011 & 0.225 $\pm$ 0.050 & 0.864 $\pm$ 0.011 \\
Qwen2.5-VL-72B-Instruct & 0.178 $\pm$ 0.053 & 0.441 $\pm$ 0.013 & 0.235 $\pm$ 0.048 & 0.865 $\pm$ 0.011 \\
\bottomrule
\end{tabular}
\label{tab:text_similarity_metrics_summary}
\vspace{-0.1in}
\end{table*}

\begin{table*}[!ht]
\centering
\footnotesize
\setlength{\tabcolsep}{3pt}
\renewcommand{\arraystretch}{0.8}
\caption{Average Image Quality Metrics (mean $\pm$ std) Across All Categories for Text-to-Image Task on Content Quality Benchmark. $\uparrow$: larger is better; $\downarrow$: smaller is better}
\begin{tabular}{lccccc}
\toprule
Model & PSNR $\uparrow$ & SSIM $\uparrow$ & MSE $\downarrow$ & LPIPS $\downarrow$ & CLIP $\uparrow$ \\
\midrule
Gemma-3-4B & 8.164 $\pm$ 0.576 & 0.347 $\pm$ 0.048 & 0.171 $\pm$ 0.018 & 0.723 $\pm$ 0.012 & 0.716 $\pm$ 0.020 \\
Gemma-3-12B & \textbf{8.516 $\pm$ 0.662} & \textbf{0.373 $\pm$ 0.049} & \textbf{0.160 $\pm$ 0.020} & \textbf{0.700 $\pm$ 0.017} & \underline{0.727 $\pm$ 0.020} \\
Gemma-3-27B & 8.184 $\pm$ 0.621 & \underline{0.349 $\pm$ 0.050} & 0.174 $\pm$ 0.022 & 0.716 $\pm$ 0.016 & 0.722 $\pm$ 0.018 \\
Qwen2.5-VL-7B-Instruct & \underline{8.260 $\pm$ 0.662} & 0.340 $\pm$ 0.056 & \underline{0.169 $\pm$ 0.022} & 0.721 $\pm$ 0.020 & 0.723 $\pm$ 0.022 \\
Qwen2.5-VL-32B-Instruct & 8.184 $\pm$ 0.572 & 0.342 $\pm$ 0.051 & 0.170 $\pm$ 0.018 & \underline{0.716 $\pm$ 0.013} & \textbf{0.727 $\pm$ 0.020} \\
Qwen2.5-VL-72B-Instruct & 7.970 $\pm$ 0.599 & 0.332 $\pm$ 0.052 & 0.179 $\pm$ 0.021 & 0.731 $\pm$ 0.013 & 0.716 $\pm$ 0.021 \\
\bottomrule
\end{tabular}
\label{tab:image_quality_metrics_summary}
\end{table*}

\begin{table*}[!ht]
\centering
\footnotesize
\setlength{\tabcolsep}{8pt}
\renewcommand{\arraystretch}{0.8}
\caption{Category-wise Average Text Quality Metrics (mean $\pm$ std) Across All Models for Image-to-Text tasks on Content Quality Benchmark. $\uparrow$: larger is better}
\renewcommand{\arraystretch}{0.9}
\begin{tabular}{lcccc}
\toprule
Category & Cosine $\uparrow$ & Euclidean $\uparrow$ & Overlap $\uparrow$ & BERTScore $\uparrow$ \\
\midrule
All Beauty & \underline{0.196 $\pm$ 0.018} & \underline{0.443 $\pm$ 0.003} & \underline{0.243 $\pm$ 0.019} & 0.865 $\pm$ 0.004 \\
Arts, Crafts \& Sewing & 0.159 $\pm$ 0.013 & 0.437 $\pm$ 0.002 & 0.216 $\pm$ 0.016 & 0.862 $\pm$ 0.003 \\
Electronics & 0.168 $\pm$ 0.015 & 0.438 $\pm$ 0.003 & 0.235 $\pm$ 0.017 & 0.865 $\pm$ 0.004 \\
Home \& Kitchen & 0.161 $\pm$ 0.015 & 0.437 $\pm$ 0.002 & 0.218 $\pm$ 0.015 & 0.862 $\pm$ 0.003 \\
Industrial \& Scientific & 0.144 $\pm$ 0.012 & 0.434 $\pm$ 0.002 & 0.199 $\pm$ 0.016 & 0.858 $\pm$ 0.004 \\
Musical Instruments & 0.156 $\pm$ 0.014 & 0.436 $\pm$ 0.002 & 0.212 $\pm$ 0.018 & 0.863 $\pm$ 0.004 \\
Office Products & 0.171 $\pm$ 0.017 & 0.439 $\pm$ 0.003 & 0.234 $\pm$ 0.019 & 0.865 $\pm$ 0.004 \\
Toys \& Games & 0.176 $\pm$ 0.017 & 0.439 $\pm$ 0.003 & 0.226 $\pm$ 0.019 & \underline{0.867 $\pm$ 0.004} \\
Video Games & \textbf{0.341 $\pm$ 0.042} & \textbf{0.490 $\pm$ 0.026} & \textbf{0.389 $\pm$ 0.050} & \textbf{0.897 $\pm$ 0.008} \\
\bottomrule
\end{tabular}
\label{tab:text_similarity_metrics_by_category_averages}
\end{table*}

\begin{table*}[!ht]
\centering
\footnotesize
\renewcommand{\arraystretch}{0.8}
\setlength{\tabcolsep}{4pt}
\caption{Category-wise Average Image Quality Metrics (mean $\pm$ std) Across All Models for Text-to-Image Tasks on Content Quality Benchmark. $\uparrow$: larger is better; $\downarrow$: smaller is better}
\begin{tabular}{lccccc}
\toprule
Category & PSNR $\uparrow$ & SSIM $\uparrow$ & MSE $\downarrow$ & LPIPS $\downarrow$ & CLIP $\uparrow$ \\
\midrule
All Beauty & 8.536 $\pm$ 0.232 & \textbf{0.422 $\pm$ 0.014} & 0.158 $\pm$ 0.007 & \underline{0.705 $\pm$ 0.015} & 0.708 $\pm$ 0.007 \\
Arts, Crafts \& Sewing & 8.758 $\pm$ 0.173 & 0.288 $\pm$ 0.011 & 0.155 $\pm$ 0.005 & 0.731 $\pm$ 0.009 & 0.698 $\pm$ 0.005 \\
Electronics & 7.697 $\pm$ 0.168 & 0.381 $\pm$ 0.016 & 0.187 $\pm$ 0.006 & 0.726 $\pm$ 0.012 & \underline{0.752 $\pm$ 0.005} \\
Home \& Kitchen & \textbf{8.997 $\pm$ 0.217} & 0.357 $\pm$ 0.015 & \textbf{0.144 $\pm$ 0.005} & 0.709 $\pm$ 0.008 & \textbf{0.756 $\pm$ 0.005} \\
Industrial \& Scientific & \underline{8.950 $\pm$ 0.270} & \underline{0.419 $\pm$ 0.012} & \underline{0.149 $\pm$ 0.009} & \textbf{0.688 $\pm$ 0.013} & 0.735 $\pm$ 0.006 \\
Musical Instruments & 7.304 $\pm$ 0.258 & 0.348 $\pm$ 0.023 & 0.203 $\pm$ 0.013 & 0.722 $\pm$ 0.015 & 0.724 $\pm$ 0.005 \\
Office Products & 8.289 $\pm$ 0.132 & 0.330 $\pm$ 0.010 & 0.168 $\pm$ 0.003 & 0.737 $\pm$ 0.007 & 0.705 $\pm$ 0.005 \\
Toys \& Games & 7.962 $\pm$ 0.238 & 0.275 $\pm$ 0.020 & 0.174 $\pm$ 0.009 & 0.718 $\pm$ 0.014 & 0.708 $\pm$ 0.006 \\
Video Games & 7.424 $\pm$ 0.078 & 0.303 $\pm$ 0.018 & 0.195 $\pm$ 0.005 & 0.725 $\pm$ 0.013 & 0.711 $\pm$ 0.009 \\
\bottomrule
\end{tabular}
\label{tab:image_quality_metrics_by_category_averages}
\end{table*}

\subsection{GRPO Training}
\label{sec:grpo_training}
To further improve the generation quality of MLLMs under missing modality conditions, we explore a reinforcement learning approach. Specifically, we utilize Group Relative Policy Optimization (GRPO)~\citep{shao2024deepseekmath}, aiming to explore aligning existing MLLMs to the modality completion tasks. We chose GRPO for its training efficiency and its strong alignment with our rule-based reward function, which measures the similarity between the generated content and the original data. We apply this approach to our defined two tasks from the Problem Formulation: \textbf{Image-to-Text} and \textbf{Text-to-Image} generation.

For each task, we design task-specific reward functions to measure the similarity between the generated and target modality outputs, thereby providing effective learning signals. In the Image-to-Text task, the MLLM completes textual descriptions from the product cover. To evaluate semantic fidelity, we compute the reward as the cosine similarity between the generated and ground-truth texts using embeddings from the ``sentence-transformers/all-MiniLM-L6-v2'' Embedding model following~\citep{ghalandari2022efficient}. In the Text-to-Image task, we employ the advanced and efficient Text-to-Image diffusion model ``stabilityai/stable-diffusion-3.5-large-turbo'' from the Hugging Face platform\footnote{https://huggingface.co/} to synthesize images based on the MLLM-completed prompt for product cover generation. The reward is then calculated using a CLIP-based~\citep{vaze2023genecis} similarity model to assess the alignment between the generated image and the intended content. For the training dataset, we sample a total of 500 items, resulting in 4,500 images across nine categories. To prevent data leakage, all sampled items are selected from outside the benchmark dataset.

\begin{figure}
    \centering
    \includegraphics[width=\linewidth]{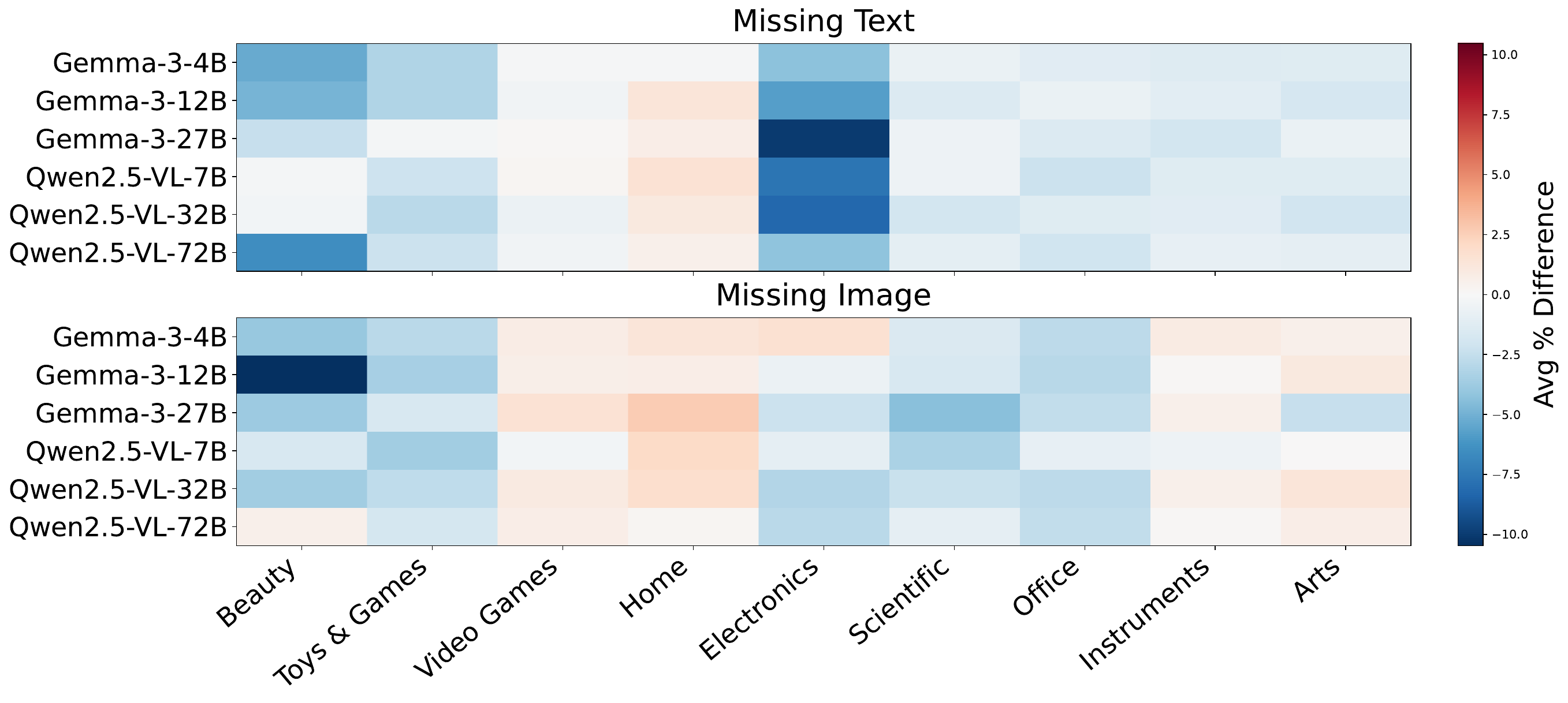}
    \caption{Heatmap Illustrating the Differences Among MLLM Completion  Methods Across Various Recommendation Categories. Colors closer to blue indicate greater performance degradation compared to using the ground-truth multimodal inputs.}
    \label{fig:heatmap_recommendation}
\end{figure}

\section{Experiments}
\subsection{Implementation Details}
We employ the MLLMs Qwen/Qwen2.5-VL-7B-Instruct, Qwen/Qwen2.5-VL-32B-Instruct, Qwen/Qwen2.5-VL-72B-Instruct, google/gemma-3-4b-it, google/gemma-3-12b-it, and google/gemma-3-27b-it from the Hugging Face platform\footnote{\url{https://huggingface.co/}} with 4-bit quantization for inference, and adopt stabilityai/stable-diffusion-3.5-large-turbo for efficient, high-quality image generation using the default 4-step generation setting. For GRPO fine-tuning, we utilize the Transformer Reinforcement Learning (TRL)\footnote{\url{https://huggingface.co/docs/trl/}} training framework following the GAD-cell/vlm-grpo implementation\footnote{\url{https://github.com/GAD-cell/vlm-grpo}}, set the LoRA rank to 16 and the learning rate in $\{1\times10^{-6}, 1\times10^{-5}, 1\times10^{-4}\}$ and select $1\times10^{-5}$ as the default learning rate. We set the group size to 16. To stabilize the training of the Text-to-Image diffusion model, for each group sample, we generate four images and average their scores. Training continues until the reward curve converges. All experiments are conducted on a single 80 GB H100 GPU. We will release our code and datasets for future research. Our code will be available to facilitate future research. 

\subsection{Benchmark Analysis}
In this section, we analyze the results of our two benchmarks: the \textbf{Completion Quality Benchmark} and the \textbf{Recommendation Benchmark}. Due to page limitations, we summarize the findings using aggregated tables and figures in this section, while the detailed results are provided in the Appendix.

\paragraph{Completion Quality Benchmark.} For the CQBench, we evaluate six advanced MLLMs from the Gemma-3 and Qwen2.5-VL families. A general observation can be shown in Tables~\ref{tab:text_similarity_metrics_summary} and~\ref{tab:image_quality_metrics_summary}, BERTScore and CLIPScore are consistently high with minimal variation across models, indicating strong semantic alignment. In contrast, word- and patch-level metrics (e.g., Cosine, Euclidean, Overlap, PSNR, SSIM, MSE, LPIPS) yield much lower score ranges, suggesting that current MLLMs capture semantic meaning well but struggle with exact word-level or pixel-level correspondence. More detailed result analysis is reported from two perspectives—model scale and modality category—aiming to reveal the capability of MLLMs in completing multi-modalities of e-commerce products.

\noindent \textit{Model Scales.}
 As shown in Figure~\ref{fig:radar_presentatoin}, within the Gemma-3 line, larger clearly means better for the \textbf{Image-to-Text} task: the 27B variant attains the highest cosine similarity ($0.210$) and BERTScore ($0.872$) across all six models, whereas the 4B and 12B models lag behind by around $20\%$ on Cosine similarity as described in Table~\ref{tab:text_similarity_metrics_summary}.  
In sharp contrast, Qwen2.5-VL peaks at its 7B scale; both the 32B and 72B versions slide by roughly $10\%$–$15\%$ on the same metrics. 
For \textbf{Text-to-Image}, scale is not monotonic: Gemma’s 12B delivers the best PSNR ($8.52$) and SSIM ($0.373$), while Qwen again favours the 7B version (PSNR $8.26$)  as described in Table~\ref{tab:image_quality_metrics_summary}. 
These opposite trends indicate that capacity alone does not dictate cross-modal quality.

\begin{table*}[ht]
\centering
\footnotesize
\setlength{\tabcolsep}{2pt}
\renewcommand{\arraystretch}{0.8}
\caption{Benchmark of MLLMs on Image-to-Text (I$\rightarrow$T) and Text-to-Image (T$\rightarrow$I).  
Best and second‐best per metric are \textbf{bold} and \underline{underlined}. ``Rel. Diff. (vs. GT)'' denotes the relative difference from the ground truth.}
\begin{tabular}{clcccccccc}
\toprule
\textbf{Task} & \textbf{Completion Models} & \textbf{R@5} & \textbf{R@10} &
\textbf{R@20} & \textbf{R@50} &
\textbf{N@5} & \textbf{N@10} & \textbf{N@20} & \textbf{N@50} \\
\midrule
 & Ground Truth & 0.0346 & 0.0552 & 0.0851 & 0.1461 & 0.0224 & 0.0291 & 0.0367 & 0.0488 \\
\midrule
\multirow{7}{*}{I$\rightarrow$T}
 & Gemma-3-4B                   & \textbf{0.0344} & 0.0547          & \textbf{0.0856} & \textbf{0.1469} & \textbf{0.0221} & 0.0287          & \textbf{0.0366} & \textbf{0.0488} \\
 & Gemma-3-12B                  & \underline{0.0342} & 0.0546        & 0.0846          & 0.1443          & \underline{0.0221} & \underline{0.0287} & 0.0364          & 0.0483          \\
 & Gemma-3-27B                  & 0.0339          & \underline{0.0547} & 0.0844      & 0.1454          & 0.0219          & 0.0287          & 0.0362          & 0.0483          \\
 & Qwen2.5-VL-7B-Instruct       & 0.0339          & 0.0546          & 0.0852          & 0.1462          & 0.0219          & 0.0286          & 0.0364          & 0.0485          \\
 & Qwen2.5-VL-32B-Instruct      & 0.0342          & 0.0545          & \underline{0.0855} & 0.1458      & 0.0220          & 0.0286          & \underline{0.0365} & 0.0485          \\
 & Qwen2.5-VL-72B-Instruct      & 0.0340          & \textbf{0.0550} & 0.0854          & \underline{0.1463} & 0.0219       & \textbf{0.0287} & 0.0365          & \underline{0.0486} \\
\cline{2-10}
 & Rel.\ Diff.\ (vs. GT)         & \(-0.58\%\)     & \(-0.36\%\)     & \(+0.58\%\)     & \(+0.55\%\)     & \(-1.34\%\)     & \(-1.37\%\)     & \(-0.27\%\)     & \(0.00\%\)      \\
\midrule
\multirow{7}{*}{T$\rightarrow$I}
 & Gemma-3-4B                   & \underline{0.0340} & \underline{0.0542} & 0.0841      & 0.1437          & \underline{0.0221} & \underline{0.0286} & \underline{0.0362} & 0.0481          \\
 & Gemma-3-12B                  & 0.0340          & 0.0542          & 0.0837          & \textbf{0.1448} & 0.0220          & 0.0286          & 0.0360          & \textbf{0.0482} \\
 & Gemma-3-27B                  & \textbf{0.0340} & 0.0538          & \underline{0.0843} & 0.1441      & \textbf{0.0221} & 0.0285          & \underline{0.0362} & 0.0482          \\
 & Qwen2.5-VL-7B-Instruct       & 0.0339          & \underline{0.0542} & \textbf{0.0844} & 0.1436      & 0.0220          & \textbf{0.0286} & \textbf{0.0363} & 0.0481          \\
 & Qwen2.5-VL-32B-Instruct      & 0.0339          & 0.0538          & 0.0840          & 0.1440          & 0.0219          & 0.0284          & 0.0361          & 0.0480          \\
 & Qwen2.5-VL-72B-Instruct      & 0.0339          & \textbf{0.0543} & 0.0842          & \underline{0.1443} & 0.0220       & 0.0286          & 0.0362          & \underline{0.0482} \\
\cline{2-10}
 & Rel.\ Diff.\ (vs. GT)         & \(-1.73\%\)     & \(-1.63\%\)     & \(-0.82\%\)     & \(-0.89\%\)     & \(-1.34\%\)     & \(-1.72\%\)     & \(-1.09\%\)     & \(-1.23\%\)     \\
\bottomrule
\end{tabular}
\vspace{-0.1in}

\label{tab:mllm_modality_benchmark}
\end{table*}

\begin{table}[!ht]
\centering
\small
\setlength{\tabcolsep}{5pt}
\renewcommand{\arraystretch}{0.8}
\caption{Average Text Quality Metrics (mean $\pm$ std) Across All Categories for GRPO Training for Qwen2.5-VL-Instruct. $\uparrow$: larger is better}
\begin{tabular}{lccccc}
\toprule
Model & Cosine $\uparrow$ & Overlap $\uparrow$ & BERTScore $\uparrow$ \\
\midrule
Before GRPO & 0.196 $\pm$ 0.061 & 0.254 $\pm$ 0.060 & 0.870 $\pm$ 0.011 \\
After GRPO & \textbf{0.249 $\pm$ 0.084} & \textbf{0.295 $\pm$ 0.085} & \textbf{0.873 $\pm$ 0.017}  \\
\bottomrule
\end{tabular}

\vspace{-0.15in}
\label{tab:grpo_text_similarity}
\end{table}

\begin{table}[!ht]
\centering
\small
\renewcommand{\arraystretch}{0.8}
\caption{Average Image Quality Metrics (mean $\pm$ std) Across All categories for GRPO Training. $\uparrow$: larger is better; $\downarrow$: smaller is better}
\setlength{\tabcolsep}{5pt}
\begin{tabular}{lccccc}
\toprule
Model & SSIM $\uparrow$ & MSE $\downarrow$ & CLIP $\uparrow$ \\
\midrule
Before GRPO & 0.340 $\pm$ 0.056& 0.169 $\pm$ 0.022 & 0.723 $\pm$ 0.022\\
After GRPO  & 0.339 $\pm$ 0.056 & 0.170 $\pm$ 0.022 & 0.723 $\pm$ 0.022 \\
\bottomrule
\end{tabular}

\vspace{-0.15in}
\label{tab:grpo_image_similarity}
\end{table}

\noindent \textit{Category-level divergence.}
Radar visualisations expose orthogonal difficulty profiles between the two tasks as described in Figure~\ref{fig:radar_presentatoin}.  
On \textbf{Image-to-Text}, \emph{Video Games} is an outlier—average cosine similarity soars to $0.341$, almost double the mean as described in Table~\ref{tab:text_similarity_metrics_by_category_averages}.  
Switching to \textbf{Text-to-Image}, gaps narrow overall, yet polarity flips: \emph{Home \& Kitchen} items yield the most similar reconstructions (PSNR $8.997$), whereas \emph{Video Games} sinks to the worst ($7.42$).  
The same product category, \emph{Video Games}, thus oscillates between the ``simplest'' and the ``hardest'' tasks. This observation validates the benchmark by showing that, even within the same category, the two tasks exhibit completely different levels of difficulty.

\paragraph{Recommendation Benchmark.} 
Across nine categories and three recommenders, replacing the ground-truth modality with generated content leads to only minor performance changes: Recall and NDCG vary by at most $2\%$, with relative deltas ranging from $-0.6\%$ to $+0.6\%$ across eight metrics. Model-wise, Gemma-3-4B unexpectedly achieves the best performance for \textbf{Image-to-Text} completion (Recall@10 $=0.0547$). For \textbf{Text-to-Image}, scores are tightly clustered, yet Qwen-7B secures three best results and one runner-up, highlighting its balanced trade-off between generation quality and efficiency. At the category level, as shown in Figure~\ref{fig:heatmap_recommendation}, \emph{Electronics} is the most challenging under the missing-text scenario. A notable outlier is Gemma-3-12B on the \emph{Beauty} dataset, where it performs worst; we attribute this to Beauty’s small dataset size, as its performance normalizes on larger categories. Compared with the Content Quality Benchmark, recommendation performance with completed modalities shows weak correlation, likely because recommendation mainly relies on semantic-level representations \citep{zhou2023tale} with minor variations across models and categories on CQBench.

\paragraph{Summary.}
Our benchmark yields four key insights. For the \textit{CQBench}, (1) current MLLMs effectively capture semantic meaning but struggle with precise word-level and pixel-level correspondence; (2) model scale exhibits non-monotonic effects across architectures and tasks, indicating that larger size alone does not guarantee better multimodal completion; (3) and completion difficulty varies substantially by task direction and product category, revealing inherent asymmetries. From the \textit{Recommendation Benchmark}, we further observe that (4) most MLLM-generated modalities achieve performance comparable to ground-truth modalities, despite very few specific cases; (5) recommendation performance with completed modalities does not strongly correlate with the Content Quality Benchmark from both model-wise and category-wise.

\subsection{GRPO Results}
Motivated by the performance gaps observed in the Completion Quality benchmark—where existing MLLMs exhibit limited alignment with task objectives, likely due to insufficient exposure during pretraining—we investigate \textit{Group Relative Policy Optimization (GRPO)} as a fine-tuning strategy to better align model behavior with both directions of the presentation track. We select \textit{Qwen2.5-VL-7B-Instruct} as the base model, given its strong performance on our benchmark and its computational efficiency. In this section, we present only three metrics due to space limitations; the remaining metrics exhibit consistent trends.

For \textit{Image-to-Text}, GRPO yields substantial improvements; for example, cosine similarity increases from $0.196$ to $0.249$ (+27\%), as shown in Table~\ref{tab:grpo_text_similarity}, with the increases of the other two metrics. BERTScore shows a relatively marginal improvement, primarily because its initial performance is already high, making further enhancement challenging. This confirms the effectiveness of GRPO in this direction. However, for \textit{Text-to-Image} as shown in Table~\ref{tab:grpo_image_similarity}, no clear gains are observed, with the reward signal showing no upward trend during training—indicating that further investigation is needed to improve alignment in this more challenging direction. 

These findings demonstrate that GRPO is highly effective for aligning models in the Image-to-Text direction, substantially improving generation quality. However, its limited impact on Text-to-Image highlights a more complex optimization landscape, suggesting the need for task-specific strategies and further exploration in this direction.

\section{Conclusion and Future Work}
We study the underexplored task of missing-modality completion in e-commerce by introducing MMPCBench, the first systematic benchmark comprising the Content Quality Benchmark (CQBench) and the Recommendation Benchmark (RecBench), covering both Image-to-Text and Text-to-Image tasks across nine product categories with six state-of-the-art MLLMs from the Qwen2.5-VL and Gemma-3 families.

Experiments on CQBench reveal that current MLLMs capture high-level semantic meaning but struggle with precise word- and pixel-level alignment. Model scaling exhibits non-monotonic effects—Qwen2.5-VL peaks at 7B parameters while Gemma-3 improves with scale—indicating that capacity alone does not guarantee better completion. Task difficulty varies substantially by direction and category, with Video Games excelling in I→T but proving most challenging for T→I. On RecBench, most completed modalities achieve recommendation performance close to ground-truth, though the correlation between content quality and recommendation accuracy remains weak, suggesting semantic alignment matters more than pixel-perfect reconstruction. Our GRPO investigation enhances I→T completion substantially but shows limited effectiveness for T→I, revealing a key weakness in current alignment techniques for this more challenging direction.

MMPCBench benefits the research community by providing a standardized, reproducible evaluation framework that enables fair model comparison. Our findings about the weak correlation between content quality and recommendation performance offer insights for designing evaluation metrics and training objectives—practitioners may prioritize semantic alignment over low-level fidelity. The category-specific patterns guide model selection for particular domains, while our open-source release facilitates extensions to other applications requiring modality completion, such as medical imaging or scientific documentation.

Future work will tackle the T→I bottleneck through multi-level reward mechanisms combining perceptual and semantic metrics, potentially using two-stage diffusion refinement. Expanding MMPCBench with fine-grained attributes, multi-image scenarios, and hybrid evaluation protocols combining automatic metrics with human assessment and online testing will better capture real-world complexity and practical utility. Exploring parameter-efficient alignment methods and transfer learning across categories could improve accessibility and reduce domain-specific training requirements for broader deployment.

\appendix

\section{Appendix}
\subsection{More Results of GRPO}
\begin{figure}
    \centering
    \begin{subfigure}[t]{0.48\linewidth}
        \centering
        \includegraphics[width=\linewidth]{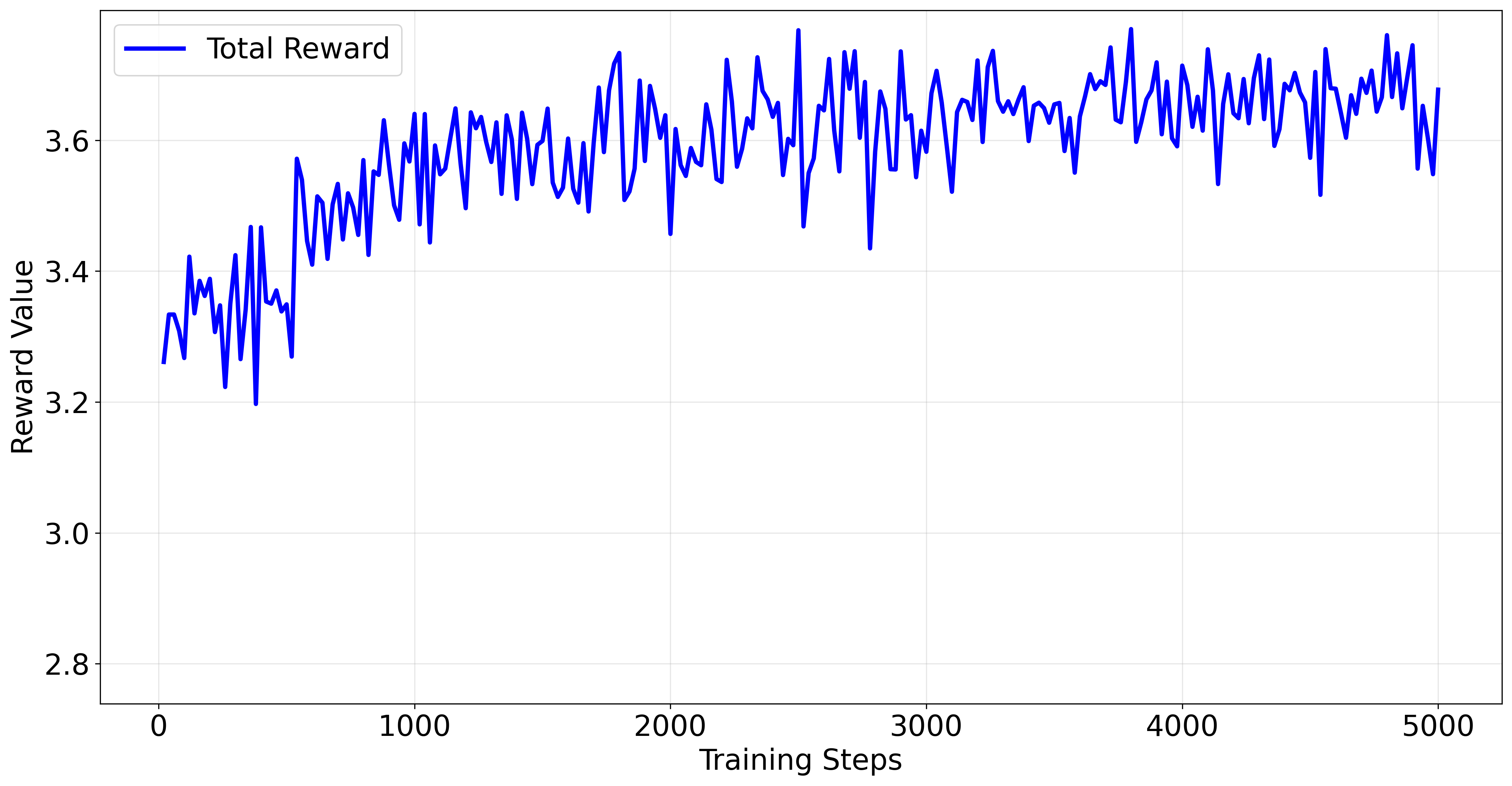}
        \caption{Image-to-Text task}
        \label{fig:i2t_grpo_curve}
    \end{subfigure}
    \hfill
    \begin{subfigure}[t]{0.48\linewidth}
        \centering
        \includegraphics[width=\linewidth]{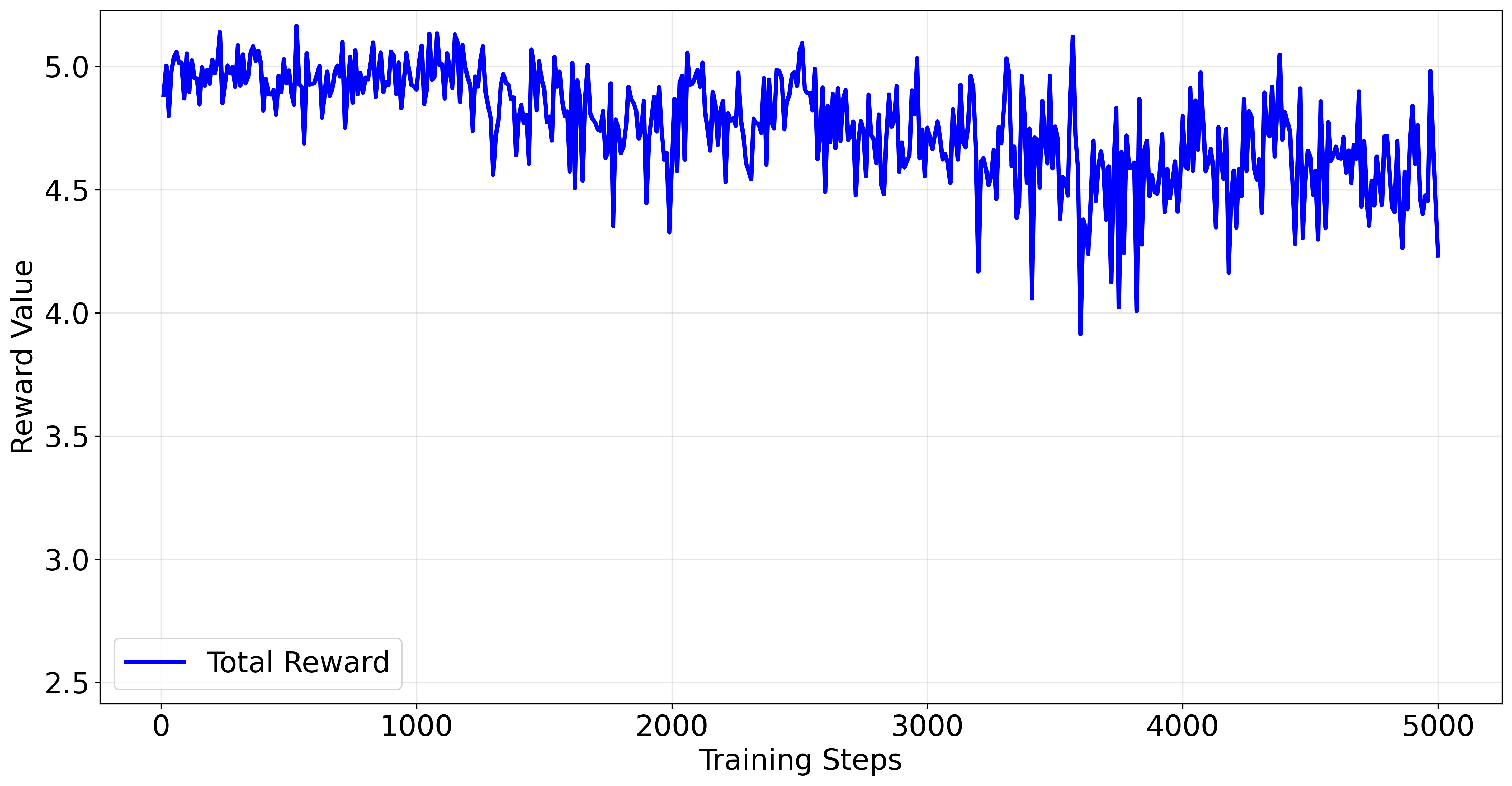}
        \caption{Text-to-Image task}
        \label{fig:t2i_grpo_curve}
    \end{subfigure}
    \caption{Reward Curves of different tasks.}
    \label{fig:grpo_curves}
\end{figure}

As shown in Figures~\ref{fig:i2t_grpo_curve} and~\ref{fig:t2i_grpo_curve}, we present the training rewards for the Image-to-Text (I$\rightarrow$T) and Text-to-Image (T$\rightarrow$I) tasks. We adopt the Qwen2.5-VL-Instruct as the default model for GRPO training. Each task incorporates an output JSON-format checking reward combined with a similarity reward. For the T$\rightarrow$I task, we additionally apply a scaling factor to adjust the reward range; we tuned and set 2 for I$\rightarrow$T, while 4 for T$\rightarrow$I, which explains the differences of absolute values.

The I$\rightarrow$T reward exhibits a clear upward trend, improving rapidly within the first 1,000 steps and then continuing to rise steadily thereafter. In contrast, the T$\rightarrow$I reward remains relatively stable during the initial 1,000 steps and subsequently begins to decline. This pattern suggests that the I$\rightarrow$T task is comparatively easier for resolving using GRPO, while the T$\rightarrow$I task may require more sophisticated reward designs to achieve consistent improvement, which should be studied in future work. 

After demonstrating the overall improvement of GRPO on the I$\rightarrow$T task, we further provide the per-category performance in Table~\ref{tab:text_similarity_grpo}. 
After GRPO training, most categories and metrics generally improve. 
For example, in the best-performing category \emph{Video Games}, the score increases from $0.37 \rightarrow 0.48$ on Cosine similarity and from $0.90 \rightarrow 0.92$ on BERTScore. 
We also observe that the performance gap among Cosine, Euclidean, and Overlap metrics is relatively large, 
whereas BERTScore exhibits a smaller improvement margin because the original model already achieves strong performance. 
In some categories (e.g., \emph{All Beauty} and \emph{Toys \& Games}), BERTScore even shows a slight drop, while, in most other categories, all metrics improve. These results indicate that GRPO-based reinforcement learning effectively strengthens the alignment between generated captions and ground-truth descriptions.

\begin{table*}
\centering
\footnotesize
\setlength{\tabcolsep}{4pt}
\renewcommand{\arraystretch}{0.6}
\caption{Text Quality Metrics (mean $\pm$ std) by Method and Category. $\uparrow$: larger is better}
\begin{tabular}{llcccc}
\toprule
Method & Category & Cosine $\uparrow$ & Euclidean $\uparrow$ & Overlap $\uparrow$ & BERTScore $\uparrow$ \\
\midrule
\multirow{9}{*}{Before GRPO} & All Beauty & \underline{0.198 $\pm$ 0.130} & \underline{0.443 $\pm$ 0.023} & 0.248 $\pm$ 0.136 & 0.866 $\pm$ 0.030 \\
 & Arts, Crafts \& Sewing & 0.165 $\pm$ 0.129 & 0.439 $\pm$ 0.033 & 0.228 $\pm$ 0.149 & 0.865 $\pm$ 0.031 \\
 & Electronics & 0.181 $\pm$ 0.118 & 0.440 $\pm$ 0.020 & 0.249 $\pm$ 0.139 & \underline{0.869 $\pm$ 0.032} \\
 & Home \& Kitchen & 0.166 $\pm$ 0.105 & 0.438 $\pm$ 0.024 & 0.224 $\pm$ 0.122 & 0.864 $\pm$ 0.027 \\
 & Industrial \& Scientific & 0.152 $\pm$ 0.117 & 0.436 $\pm$ 0.019 & 0.208 $\pm$ 0.139 & 0.860 $\pm$ 0.034 \\
 & Musical Instruments & 0.163 $\pm$ 0.127 & 0.438 $\pm$ 0.022 & 0.225 $\pm$ 0.148 & 0.866 $\pm$ 0.032 \\
 & Office Products & 0.189 $\pm$ 0.142 & 0.443 $\pm$ 0.039 & \underline{0.252 $\pm$ 0.168} & 0.869 $\pm$ 0.034 \\
 & Toys \& Games & 0.184 $\pm$ 0.131 & 0.441 $\pm$ 0.024 & 0.235 $\pm$ 0.138 & 0.870 $\pm$ 0.031 \\
 & Video Games & \textbf{0.365 $\pm$ 0.255} & \textbf{0.492 $\pm$ 0.101} & \textbf{0.420 $\pm$ 0.248} & \textbf{0.901 $\pm$ 0.043} \\
 \midrule
\multirow{9}{*}{After GRPO} & All Beauty & \underline{0.270 $\pm$ 0.140} & \underline{0.455 $\pm$ 0.026} & 0.263 $\pm$ 0.121 & 0.857 $\pm$ 0.024 \\
 & Arts, Crafts \& Sewing & 0.202 $\pm$ 0.143 & 0.445 $\pm$ 0.032 & 0.269 $\pm$ 0.160 & 0.870 $\pm$ 0.033 \\
 & Electronics & 0.233 $\pm$ 0.136 & 0.449 $\pm$ 0.030 & \underline{0.300 $\pm$ 0.150} & \underline{0.880 $\pm$ 0.034} \\
 & Home \& Kitchen & 0.205 $\pm$ 0.119 & 0.444 $\pm$ 0.027 & 0.268 $\pm$ 0.141 & 0.870 $\pm$ 0.029 \\
 & Industrial \& Scientific & 0.190 $\pm$ 0.135 & 0.443 $\pm$ 0.034 & 0.251 $\pm$ 0.152 & 0.867 $\pm$ 0.033 \\
 & Musical Instruments & 0.195 $\pm$ 0.144 & 0.444 $\pm$ 0.028 & 0.255 $\pm$ 0.168 & 0.869 $\pm$ 0.036 \\
 & Office Products & 0.226 $\pm$ 0.150 & 0.450 $\pm$ 0.044 & 0.292 $\pm$ 0.174 & 0.876 $\pm$ 0.035 \\
 & Toys \& Games & 0.244 $\pm$ 0.132 & 0.451 $\pm$ 0.024 & 0.230 $\pm$ 0.115 & 0.854 $\pm$ 0.024 \\
 & Video Games & \textbf{0.477 $\pm$ 0.318} & \textbf{0.575 $\pm$ 0.208} & \textbf{0.528 $\pm$ 0.301} & \textbf{0.916 $\pm$ 0.052} \\
\bottomrule
\end{tabular}
\label{tab:text_similarity_grpo}
\end{table*}

\subsection{More Analysis of the Results}

\noindent \textbf{(1) CQBench: Image-to-Text Task}.
From Table~\ref{tab:CQBench_text_ap} it is clear that the Video Games category is the best for Image-to-Text generation: across every model and metric it posts the highest scores.  For example, with Gemma-3-27B the results reach Cosine 0.394 ± 0.292, Euclidean 0.530 ± 0.175, Overlap 0.446 ± 0.283, and BERTScore 0.906 ± 0.051. 

Looking at the model families, Gemma-3 consistently outperforms Qwen2.5-VL.  Both the 27B and 12B checkpoints of Gemma-3 score higher across almost all datasets than any of the three Qwen2.5-VL variants.  Within families the scaling behaviour diverges: Gemma-3 improves steadily as its parameter count grows, whereas Qwen2.5-VL peaks at the 7B model, followed by the 72B version, with the 32B model trailing behind.

\noindent \textbf{(2) CQBench: Text-to-Image Task}.
From the Table~\ref{tab:CQBench_image_ap}, we can see ``Home \& Kitchen" and ``Industrial \& Scientific" datasets have the best result in the T$\rightarrow$I task. Across models, Gemma-3-12B is the strongest overall. In the ``All Beauty" dataset, Gemma-3-12B hits the top SSIM 0.446 ± 0.181, lowest MSE 0.148 ± 0.073, and lowest LPIPS 0.684 ± 0.117. Similar to Table~\ref{tab:CQBench_text_ap}, scaling helps Gemma, but not Qwen2.5-VL. Gemma's metrics improve steadily from 4B → 12B → 27B. While Qwen2.5-VL: the 7B model actually outperforms the 32B and 72B variants on PSNR, SSIM, and LPIPS.

\noindent \textbf{(3) RecBench: Image-to-Text Task}.
According to Tables~\ref{tab:performance_vbpr_i_to_t}, \ref{tab:performance_bm3_i_to_t}, and \ref{tab:performance_freedom_i_to_t}, the captions generated by multimodal large-language models (MLLMs) achieve nearly the same performance as the original texts on all three recommenders. In VBPR, BM3, and FREEDOM, the evaluation scores for MLLM-completed captions remain very close to the real-text baselines. For example, in the FREEDOM model for the ``Arts" dataset, the ground-truth Recall@20 is 0.1446, while the caption created by Gemma-3-27B posts an almost identical 0.1442. And for the ``Video Games" category under BM3, the ground-truth NDCG@50 is 0.0814—the exact same value achieved by Qwen2.5-VL-7B. This consistency shows that the generated captions successfully preserve the core semantic signals required by these different recommendation algorithms. 

Unlike the patterns in Tables~\ref{tab:CQBench_text_ap} and ~\ref{tab:CQBench_image_ap}, no single MLLM is universally superior for this task; the optimal model depends strongly on the product category and the recommender being used. With the FREEDOM, for instance, Qwen2.5-VL-7B and Qwen2.5-VL-72B perform best in the  ``Instruments" category (NDCG@20 = 0.0665), whereas Gemma-3-27B excels in  ``Toys \& Games" (NDCG@20 = 0.0382). These variations highlight that, although generated captions can effectively replace missing modality information, the best generator is context-specific.

\noindent \textbf{(4) RecBench: Text-to-Image Task}.
According to Tables~\ref{tab:performance_vbpr_t_to_i}, \ref{tab:performance_bm3_t_to_i}, and \ref{tab:performance_freedom_t_to_i}, the Text-to-Image task shows that replacing ground-truth images with MLLM-generated ones has minimal impact on the recommendation performance of VBPR, BM3, and FREEDOM. The Recall and NDCG metrics for completions are consistently on par with the ground-truth baseline. For example, using the FREEDOM recommender on the ``Arts" dataset, the ground-truth Recall@10 is 0.0912, while the Qwen2.5-VL-32B model not only matches but slightly exceeds this with a score of 0.0926. Similarly, for the BM3 model in the ``Home" category, the ground-truth NDCG@50 is 0.0473, and Gemma-3-12B achieves a higher score of 0.0494. This demonstrates that the generated images are effective surrogates for original images in recommendation tasks.

\begin{table*}
\centering
\renewcommand{\arraystretch}{0.6}
\footnotesize
\setlength{\tabcolsep}{4pt}
\caption{Text Quality Metrics (mean $\pm$ std) by Model and Category. $\uparrow$: larger is better}
\begin{tabular}{llcccc}
\toprule
Model & Category & Cosine $\uparrow$ & Euclidean $\uparrow$ & Overlap $\uparrow$ & BERTScore $\uparrow$ \\
\midrule
\multirow{9}{*}{Gemma-3-4B} & All Beauty & 0.186 $\pm$ 0.111 & 0.441 $\pm$ 0.019 & 0.224 $\pm$ 0.113 & 0.862 $\pm$ 0.025 \\
 & Arts, Crafts \& Sewing & 0.149 $\pm$ 0.113 & 0.435 $\pm$ 0.020 & 0.195 $\pm$ 0.121 & 0.857 $\pm$ 0.027 \\
 & Electronics & 0.155 $\pm$ 0.097 & 0.436 $\pm$ 0.016 & 0.217 $\pm$ 0.113 & 0.862 $\pm$ 0.028 \\
 & Home \& Kitchen & 0.149 $\pm$ 0.087 & 0.435 $\pm$ 0.013 & 0.200 $\pm$ 0.102 & 0.859 $\pm$ 0.024 \\
 & Industrial \& Scientific & 0.129 $\pm$ 0.097 & 0.432 $\pm$ 0.015 & 0.176 $\pm$ 0.111 & 0.854 $\pm$ 0.026 \\
 & Musical Instruments & 0.137 $\pm$ 0.105 & 0.433 $\pm$ 0.017 & 0.185 $\pm$ 0.118 & 0.857 $\pm$ 0.028 \\
 & Office Products & 0.149 $\pm$ 0.102 & 0.435 $\pm$ 0.017 & 0.205 $\pm$ 0.120 & 0.859 $\pm$ 0.027 \\
 & Toys \& Games & 0.164 $\pm$ 0.103 & 0.437 $\pm$ 0.017 & 0.207 $\pm$ 0.111 & 0.863 $\pm$ 0.025 \\
 & Video Games & 0.279 $\pm$ 0.187 & 0.460 $\pm$ 0.043 & 0.312 $\pm$ 0.174 & 0.883 $\pm$ 0.035 \\
\cmidrule{2-6}
\multirow{9}{*}{Gemma-3-12B} & All Beauty & 0.208 $\pm$ 0.120 & 0.444 $\pm$ 0.021 & 0.250 $\pm$ 0.124 & 0.867 $\pm$ 0.028 \\
 & Arts, Crafts \& Sewing & 0.168 $\pm$ 0.122 & 0.438 $\pm$ 0.021 & 0.222 $\pm$ 0.136 & 0.863 $\pm$ 0.029 \\
 & Electronics & 0.178 $\pm$ 0.114 & 0.440 $\pm$ 0.020 & 0.241 $\pm$ 0.128 & 0.866 $\pm$ 0.030 \\
 & Home \& Kitchen & 0.172 $\pm$ 0.097 & 0.438 $\pm$ 0.015 & 0.227 $\pm$ 0.112 & 0.864 $\pm$ 0.025 \\
 & Industrial \& Scientific & 0.152 $\pm$ 0.106 & 0.435 $\pm$ 0.017 & 0.205 $\pm$ 0.122 & 0.860 $\pm$ 0.028 \\
 & Musical Instruments & 0.163 $\pm$ 0.120 & 0.438 $\pm$ 0.021 & 0.218 $\pm$ 0.136 & 0.864 $\pm$ 0.030 \\
 & Office Products & 0.172 $\pm$ 0.118 & 0.439 $\pm$ 0.020 & 0.232 $\pm$ 0.142 & 0.864 $\pm$ 0.029 \\
 & Toys \& Games & 0.187 $\pm$ 0.117 & 0.441 $\pm$ 0.020 & 0.235 $\pm$ 0.129 & 0.869 $\pm$ 0.027 \\
 & Video Games & \underline{0.367 $\pm$ 0.272} & \underline{0.508 $\pm$ 0.146} & \underline{0.425 $\pm$ 0.268} & \underline{0.902 $\pm$ 0.048} \\
\cmidrule{2-6}
\multirow{9}{*}{Gemma-3-27B} & All Beauty & 0.223 $\pm$ 0.120 & 0.447 $\pm$ 0.021 & 0.273 $\pm$ 0.129 & 0.871 $\pm$ 0.027 \\
 & Arts, Crafts \& Sewing & 0.178 $\pm$ 0.121 & 0.440 $\pm$ 0.021 & 0.237 $\pm$ 0.139 & 0.866 $\pm$ 0.029 \\
 & Electronics & 0.185 $\pm$ 0.109 & 0.441 $\pm$ 0.018 & 0.259 $\pm$ 0.130 & 0.870 $\pm$ 0.031 \\
 & Home \& Kitchen & 0.183 $\pm$ 0.103 & 0.440 $\pm$ 0.017 & 0.239 $\pm$ 0.118 & 0.867 $\pm$ 0.026 \\
 & Industrial \& Scientific & 0.159 $\pm$ 0.109 & 0.437 $\pm$ 0.017 & 0.219 $\pm$ 0.131 & 0.863 $\pm$ 0.029 \\
 & Musical Instruments & 0.174 $\pm$ 0.122 & 0.439 $\pm$ 0.021 & 0.234 $\pm$ 0.146 & 0.866 $\pm$ 0.032 \\
 & Office Products & 0.190 $\pm$ 0.121 & 0.442 $\pm$ 0.028 & 0.254 $\pm$ 0.141 & 0.870 $\pm$ 0.030 \\
 & Toys \& Games & 0.201 $\pm$ 0.120 & 0.443 $\pm$ 0.022 & 0.254 $\pm$ 0.133 & 0.872 $\pm$ 0.028 \\
 & Video Games & \textbf{0.394 $\pm$ 0.292} & \textbf{0.530 $\pm$ 0.175} & \textbf{0.446 $\pm$ 0.283} & \textbf{0.906 $\pm$ 0.051} \\
\cmidrule{2-6}
\multirow{9}{*}{\begin{tabular}[c]{@{}c@{}}Qwen2.5-VL-\\Instruct\end{tabular}} & All Beauty & 0.198 $\pm$ 0.130 & 0.443 $\pm$ 0.023 & 0.248 $\pm$ 0.136 & 0.866 $\pm$ 0.030 \\
 & Arts, Crafts \& Sewing & 0.165 $\pm$ 0.129 & 0.439 $\pm$ 0.033 & 0.228 $\pm$ 0.149 & 0.865 $\pm$ 0.031 \\
 & Electronics & 0.181 $\pm$ 0.118 & 0.440 $\pm$ 0.020 & 0.249 $\pm$ 0.139 & 0.869 $\pm$ 0.032 \\
 & Home \& Kitchen & 0.166 $\pm$ 0.105 & 0.438 $\pm$ 0.024 & 0.224 $\pm$ 0.122 & 0.864 $\pm$ 0.027 \\
 & Industrial \& Scientific & 0.152 $\pm$ 0.117 & 0.436 $\pm$ 0.019 & 0.208 $\pm$ 0.139 & 0.860 $\pm$ 0.034 \\
 & Musical Instruments & 0.163 $\pm$ 0.127 & 0.438 $\pm$ 0.022 & 0.225 $\pm$ 0.148 & 0.866 $\pm$ 0.032 \\
 & Office Products & 0.189 $\pm$ 0.142 & 0.443 $\pm$ 0.039 & 0.252 $\pm$ 0.168 & 0.869 $\pm$ 0.034 \\
 & Toys \& Games & 0.184 $\pm$ 0.131 & 0.441 $\pm$ 0.024 & 0.235 $\pm$ 0.138 & 0.870 $\pm$ 0.031 \\
 & Video Games & 0.365 $\pm$ 0.255 & 0.492 $\pm$ 0.101 & 0.420 $\pm$ 0.248 & 0.901 $\pm$ 0.043 \\
\cmidrule{2-6}
\multirow{9}{*}{\begin{tabular}[c]{@{}c@{}}Qwen2.5-VL-\\32B-Instruct\end{tabular}} & All Beauty & 0.174 $\pm$ 0.113 & 0.439 $\pm$ 0.019 & 0.223 $\pm$ 0.119 & 0.860 $\pm$ 0.028 \\
 & Arts, Crafts \& Sewing & 0.143 $\pm$ 0.109 & 0.434 $\pm$ 0.017 & 0.202 $\pm$ 0.125 & 0.859 $\pm$ 0.028 \\
 & Electronics & 0.151 $\pm$ 0.098 & 0.435 $\pm$ 0.016 & 0.220 $\pm$ 0.120 & 0.863 $\pm$ 0.031 \\
 & Home \& Kitchen & 0.144 $\pm$ 0.087 & 0.434 $\pm$ 0.013 & 0.204 $\pm$ 0.105 & 0.860 $\pm$ 0.025 \\
 & Industrial \& Scientific & 0.132 $\pm$ 0.097 & 0.432 $\pm$ 0.015 & 0.186 $\pm$ 0.121 & 0.855 $\pm$ 0.029 \\
 & Musical Instruments & 0.143 $\pm$ 0.114 & 0.434 $\pm$ 0.018 & 0.200 $\pm$ 0.136 & 0.861 $\pm$ 0.031 \\
 & Office Products & 0.156 $\pm$ 0.109 & 0.436 $\pm$ 0.018 & 0.220 $\pm$ 0.131 & 0.863 $\pm$ 0.029 \\
 & Toys \& Games & 0.159 $\pm$ 0.110 & 0.437 $\pm$ 0.018 & 0.210 $\pm$ 0.120 & 0.864 $\pm$ 0.028 \\
 & Video Games & 0.316 $\pm$ 0.222 & 0.471 $\pm$ 0.061 & 0.363 $\pm$ 0.220 & 0.894 $\pm$ 0.043 \\
\cmidrule{2-6}
\multirow{9}{*}{\begin{tabular}[c]{@{}c@{}}Qwen2.5-VL-\\72B-Instruct\end{tabular}} & All Beauty & 0.184 $\pm$ 0.116 & 0.441 $\pm$ 0.021 & 0.239 $\pm$ 0.125 & 0.863 $\pm$ 0.028 \\
 & Arts, Crafts \& Sewing & 0.151 $\pm$ 0.113 & 0.436 $\pm$ 0.019 & 0.211 $\pm$ 0.128 & 0.860 $\pm$ 0.029 \\
 & Electronics & 0.155 $\pm$ 0.101 & 0.436 $\pm$ 0.016 & 0.224 $\pm$ 0.120 & 0.862 $\pm$ 0.030 \\
 & Home \& Kitchen & 0.152 $\pm$ 0.094 & 0.435 $\pm$ 0.015 & 0.216 $\pm$ 0.113 & 0.860 $\pm$ 0.026 \\
 & Industrial \& Scientific & 0.143 $\pm$ 0.102 & 0.434 $\pm$ 0.016 & 0.199 $\pm$ 0.124 & 0.856 $\pm$ 0.030 \\
 & Musical Instruments & 0.154 $\pm$ 0.121 & 0.436 $\pm$ 0.021 & 0.211 $\pm$ 0.141 & 0.863 $\pm$ 0.031 \\
 & Office Products & 0.169 $\pm$ 0.120 & 0.439 $\pm$ 0.021 & 0.238 $\pm$ 0.146 & 0.865 $\pm$ 0.032 \\
 & Toys \& Games & 0.163 $\pm$ 0.117 & 0.437 $\pm$ 0.020 & 0.213 $\pm$ 0.125 & 0.864 $\pm$ 0.028 \\
 & Video Games & 0.325 $\pm$ 0.238 & 0.478 $\pm$ 0.084 & 0.367 $\pm$ 0.230 & 0.895 $\pm$ 0.043 \\
\bottomrule
\end{tabular}
\label{tab:CQBench_text_ap}
\end{table*}

\begin{table*}
\centering
\scriptsize
\caption{Image Quality Metrics (mean $\pm$ std) by Model and Category. $\uparrow$: larger is better; $\downarrow$: smaller is better}
\renewcommand{\arraystretch}{0.7}
\setlength{\tabcolsep}{2.5pt}
\begin{tabular}{llccccc}
\toprule
Model & Category & PSNR $\uparrow$ & SSIM $\uparrow$ & MSE $\downarrow$ & LPIPS $\downarrow$ & CLIP $\uparrow$ \\
\midrule
\multirow{9}{*}{Gemma-3-4B} & All Beauty & 8.266 $\pm$ 2.141 & 0.410 $\pm$ 0.172 & 0.166 $\pm$ 0.077 & 0.723 $\pm$ 0.110 & 0.697 $\pm$ 0.102 \\
 & Arts, Crafts \& Sewing & 8.677 $\pm$ 2.454 & 0.283 $\pm$ 0.174 & 0.156 $\pm$ 0.079 & 0.738 $\pm$ 0.097 & 0.692 $\pm$ 0.106 \\
 & Electronics & 7.674 $\pm$ 1.945 & 0.382 $\pm$ 0.153 & 0.186 $\pm$ 0.073 & 0.732 $\pm$ 0.095 & 0.748 $\pm$ 0.093 \\
 & Home \& Kitchen & 9.040 $\pm$ 2.462 & 0.361 $\pm$ 0.156 & 0.143 $\pm$ 0.072 & 0.714 $\pm$ 0.090 & 0.749 $\pm$ 0.094 \\
 & Industrial \& Scientific & 8.838 $\pm$ 2.540 & 0.413 $\pm$ 0.181 & 0.152 $\pm$ 0.083 & 0.698 $\pm$ 0.114 & 0.728 $\pm$ 0.108 \\
 & Musical Instruments & 7.355 $\pm$ 1.820 & 0.352 $\pm$ 0.160 & 0.199 $\pm$ 0.077 & 0.721 $\pm$ 0.097 & 0.721 $\pm$ 0.113 \\
 & Office Products & 8.275 $\pm$ 2.277 & 0.331 $\pm$ 0.170 & 0.168 $\pm$ 0.079 & 0.740 $\pm$ 0.099 & 0.702 $\pm$ 0.111 \\
 & Toys \& Games & 7.912 $\pm$ 1.770 & 0.276 $\pm$ 0.142 & 0.176 $\pm$ 0.074 & 0.718 $\pm$ 0.073 & 0.702 $\pm$ 0.107 \\
 & Video Games & 7.444 $\pm$ 1.598 & 0.309 $\pm$ 0.168 & 0.192 $\pm$ 0.070 & 0.723 $\pm$ 0.089 & 0.711 $\pm$ 0.127 \\
\cmidrule{2-7}
\multirow{9}{*}{Gemma-3-12B} & All Beauty & 8.869 $\pm$ 2.313 & \textbf{0.446 $\pm$ 0.181} & 0.148 $\pm$ 0.073 & 0.684 $\pm$ 0.117 & 0.714 $\pm$ 0.096 \\
 & Arts, Crafts \& Sewing & 9.060 $\pm$ 2.633 & 0.309 $\pm$ 0.179 & 0.146 $\pm$ 0.081 & 0.716 $\pm$ 0.094 & 0.704 $\pm$ 0.103 \\
 & Electronics & 7.956 $\pm$ 2.148 & 0.408 $\pm$ 0.154 & 0.178 $\pm$ 0.075 & 0.708 $\pm$ 0.093 & 0.758 $\pm$ 0.089 \\
 & Home \& Kitchen & \textbf{9.384 $\pm$ 2.561} & 0.386 $\pm$ 0.158 & \textbf{0.135 $\pm$ 0.073} & 0.694 $\pm$ 0.092 & \textbf{0.761 $\pm$ 0.093} \\
 & Industrial \& Scientific & \underline{9.331 $\pm$ 2.613} & \underline{0.439 $\pm$ 0.179} & \underline{0.138 $\pm$ 0.079} & \textbf{0.669 $\pm$ 0.113} & 0.738 $\pm$ 0.103 \\
 & Musical Instruments & 7.700 $\pm$ 1.866 & 0.385 $\pm$ 0.155 & 0.184 $\pm$ 0.072 & 0.697 $\pm$ 0.094 & 0.728 $\pm$ 0.107 \\
 & Office Products & 8.481 $\pm$ 2.501 & 0.345 $\pm$ 0.175 & 0.163 $\pm$ 0.081 & 0.729 $\pm$ 0.104 & 0.707 $\pm$ 0.111 \\
 & Toys \& Games & 8.406 $\pm$ 1.778 & 0.312 $\pm$ 0.152 & 0.156 $\pm$ 0.064 & 0.694 $\pm$ 0.076 & 0.716 $\pm$ 0.108 \\
 & Video Games & 7.460 $\pm$ 1.747 & 0.330 $\pm$ 0.169 & 0.193 $\pm$ 0.074 & 0.713 $\pm$ 0.095 & 0.714 $\pm$ 0.129 \\
\cmidrule{2-7}
\multirow{9}{*}{Gemma-3-27B} & All Beauty & 8.639 $\pm$ 2.273 & 0.430 $\pm$ 0.180 & 0.156 $\pm$ 0.082 & 0.694 $\pm$ 0.110 & 0.714 $\pm$ 0.092 \\
 & Arts, Crafts \& Sewing & 8.726 $\pm$ 2.607 & 0.288 $\pm$ 0.175 & 0.158 $\pm$ 0.087 & 0.730 $\pm$ 0.095 & 0.698 $\pm$ 0.104 \\
 & Electronics & 7.729 $\pm$ 2.225 & 0.388 $\pm$ 0.158 & 0.189 $\pm$ 0.085 & 0.721 $\pm$ 0.095 & 0.751 $\pm$ 0.092 \\
 & Home \& Kitchen & 8.903 $\pm$ 2.366 & 0.354 $\pm$ 0.152 & 0.147 $\pm$ 0.074 & 0.710 $\pm$ 0.086 & 0.753 $\pm$ 0.093 \\
 & Industrial \& Scientific & 8.854 $\pm$ 2.634 & 0.413 $\pm$ 0.184 & 0.154 $\pm$ 0.089 & 0.688 $\pm$ 0.115 & 0.733 $\pm$ 0.101 \\
 & Musical Instruments & 7.126 $\pm$ 2.042 & 0.342 $\pm$ 0.157 & 0.215 $\pm$ 0.096 & 0.730 $\pm$ 0.094 & 0.720 $\pm$ 0.110 \\
 & Office Products & 8.297 $\pm$ 2.395 & 0.337 $\pm$ 0.171 & 0.169 $\pm$ 0.083 & 0.734 $\pm$ 0.097 & 0.705 $\pm$ 0.106 \\
 & Toys \& Games & 7.996 $\pm$ 1.995 & 0.279 $\pm$ 0.150 & 0.176 $\pm$ 0.084 & 0.711 $\pm$ 0.081 & 0.710 $\pm$ 0.103 \\
 & Video Games & 7.387 $\pm$ 1.861 & 0.310 $\pm$ 0.168 & 0.199 $\pm$ 0.086 & 0.727 $\pm$ 0.092 & 0.716 $\pm$ 0.129 \\
\cmidrule{2-7}
\multirow{9}{*}{\begin{tabular}[c]{@{}c@{}}Qwen2.5-VL-\\Instruct\end{tabular}} & All Beauty & 8.691 $\pm$ 2.279 & 0.419 $\pm$ 0.177 & 0.153 $\pm$ 0.075 & 0.702 $\pm$ 0.108 & 0.715 $\pm$ 0.098 \\
 & Arts, Crafts \& Sewing & 8.767 $\pm$ 2.671 & 0.279 $\pm$ 0.177 & 0.155 $\pm$ 0.079 & 0.736 $\pm$ 0.094 & 0.698 $\pm$ 0.104 \\
 & Electronics & 7.766 $\pm$ 2.163 & 0.375 $\pm$ 0.158 & 0.186 $\pm$ 0.078 & 0.723 $\pm$ 0.098 & 0.755 $\pm$ 0.089 \\
 & Home \& Kitchen & 8.981 $\pm$ 2.354 & 0.353 $\pm$ 0.153 & 0.144 $\pm$ 0.068 & 0.709 $\pm$ 0.092 & \underline{0.760 $\pm$ 0.096} \\
 & Industrial \& Scientific & 9.203 $\pm$ 2.681 & 0.428 $\pm$ 0.184 & 0.141 $\pm$ 0.075 & \underline{0.677 $\pm$ 0.122} & 0.740 $\pm$ 0.102 \\
 & Musical Instruments & 7.227 $\pm$ 1.905 & 0.318 $\pm$ 0.169 & 0.207 $\pm$ 0.088 & 0.734 $\pm$ 0.096 & 0.723 $\pm$ 0.113 \\
 & Office Products & 8.335 $\pm$ 2.354 & 0.329 $\pm$ 0.174 & 0.167 $\pm$ 0.080 & 0.735 $\pm$ 0.100 & 0.710 $\pm$ 0.110 \\
 & Toys \& Games & 7.937 $\pm$ 1.838 & 0.269 $\pm$ 0.147 & 0.175 $\pm$ 0.076 & 0.730 $\pm$ 0.082 & 0.708 $\pm$ 0.105 \\
 & Video Games & 7.437 $\pm$ 1.743 & 0.286 $\pm$ 0.164 & 0.194 $\pm$ 0.076 & 0.746 $\pm$ 0.094 & 0.697 $\pm$ 0.131 \\
\cmidrule{2-7}
\multirow{9}{*}{\begin{tabular}[c]{@{}c@{}}Qwen2.5-VL-\\32B-Instruct\end{tabular}} & All Beauty & 8.400 $\pm$ 2.059 & 0.412 $\pm$ 0.172 & 0.161 $\pm$ 0.074 & 0.713 $\pm$ 0.109 & 0.707 $\pm$ 0.097 \\
 & Arts, Crafts \& Sewing & 8.787 $\pm$ 2.558 & 0.289 $\pm$ 0.174 & 0.154 $\pm$ 0.080 & 0.726 $\pm$ 0.097 & 0.701 $\pm$ 0.105 \\
 & Electronics & 7.602 $\pm$ 1.925 & 0.372 $\pm$ 0.150 & 0.189 $\pm$ 0.072 & 0.726 $\pm$ 0.095 & 0.754 $\pm$ 0.094 \\
 & Home \& Kitchen & 8.942 $\pm$ 2.349 & 0.348 $\pm$ 0.150 & 0.145 $\pm$ 0.070 & 0.708 $\pm$ 0.087 & 0.760 $\pm$ 0.094 \\
 & Industrial \& Scientific & 8.882 $\pm$ 2.541 & 0.415 $\pm$ 0.176 & 0.150 $\pm$ 0.079 & 0.688 $\pm$ 0.114 & 0.741 $\pm$ 0.102 \\
 & Musical Instruments & 7.452 $\pm$ 1.853 & 0.360 $\pm$ 0.159 & 0.195 $\pm$ 0.075 & 0.711 $\pm$ 0.094 & 0.732 $\pm$ 0.109 \\
 & Office Products & 8.274 $\pm$ 2.265 & 0.323 $\pm$ 0.171 & 0.167 $\pm$ 0.074 & 0.736 $\pm$ 0.101 & 0.710 $\pm$ 0.109 \\
 & Toys \& Games & 7.791 $\pm$ 1.770 & 0.257 $\pm$ 0.142 & 0.180 $\pm$ 0.071 & 0.725 $\pm$ 0.077 & 0.712 $\pm$ 0.105 \\
 & Video Games & 7.525 $\pm$ 1.551 & 0.302 $\pm$ 0.173 & 0.188 $\pm$ 0.065 & 0.710 $\pm$ 0.091 & 0.723 $\pm$ 0.123 \\
\cmidrule{2-7}
\multirow{9}{*}{\begin{tabular}[c]{@{}c@{}}Qwen2.5-VL-\\72B-Instruct\end{tabular}} & All Beauty & 8.354 $\pm$ 2.132 & 0.413 $\pm$ 0.171 & 0.163 $\pm$ 0.075 & 0.718 $\pm$ 0.116 & 0.703 $\pm$ 0.101 \\
 & Arts, Crafts \& Sewing & 8.533 $\pm$ 2.449 & 0.279 $\pm$ 0.171 & 0.162 $\pm$ 0.085 & 0.739 $\pm$ 0.091 & 0.694 $\pm$ 0.109 \\
 & Electronics & 7.456 $\pm$ 1.939 & 0.362 $\pm$ 0.149 & 0.197 $\pm$ 0.085 & 0.743 $\pm$ 0.089 & 0.746 $\pm$ 0.092 \\
 & Home \& Kitchen & 8.730 $\pm$ 2.228 & 0.342 $\pm$ 0.149 & 0.151 $\pm$ 0.071 & 0.720 $\pm$ 0.083 & 0.753 $\pm$ 0.096 \\
 & Industrial \& Scientific & 8.590 $\pm$ 2.592 & 0.406 $\pm$ 0.180 & 0.162 $\pm$ 0.090 & 0.706 $\pm$ 0.120 & 0.728 $\pm$ 0.107 \\
 & Musical Instruments & 6.964 $\pm$ 1.789 & 0.334 $\pm$ 0.152 & 0.219 $\pm$ 0.094 & 0.737 $\pm$ 0.090 & 0.722 $\pm$ 0.110 \\
 & Office Products & 8.072 $\pm$ 2.101 & 0.317 $\pm$ 0.168 & 0.173 $\pm$ 0.074 & 0.749 $\pm$ 0.095 & 0.696 $\pm$ 0.112 \\
 & Toys \& Games & 7.733 $\pm$ 1.742 & 0.259 $\pm$ 0.138 & 0.182 $\pm$ 0.074 & 0.731 $\pm$ 0.071 & 0.703 $\pm$ 0.107 \\
 & Video Games & 7.293 $\pm$ 1.770 & 0.281 $\pm$ 0.165 & 0.202 $\pm$ 0.083 & 0.732 $\pm$ 0.083 & 0.705 $\pm$ 0.129 \\
\bottomrule
\end{tabular}
\label{tab:CQBench_image_ap}
\end{table*}

\begin{table*}
\centering
\scriptsize
\caption{Performance Comparison for Model VBPR on Task Image-to-Text. R and N represent Recall and NDCG.}
\renewcommand{\arraystretch}{0.6}
\setlength{\tabcolsep}{5pt}
\begin{tabular}{@{}lc|cccccccc@{}}
\toprule
Dataset & Completion Method & R@5 & N@5 & R@10 & N@10 & R@20 & N@20 & R@50 & N@50 \\
\midrule
\multirow{7}{*}{Arts} & Gemma-3-4B & \textbf{0.0162} & \textbf{0.0103} & 0.0242 & 0.0129 & 0.0406 & \underline{0.0170} & 0.0757 & 0.0239 \\
& Gemma-3-12B & 0.0156 & 0.0100 & 0.0243 & 0.0128 & 0.0401 & 0.0168 & 0.0768 & 0.0241 \\
& Gemma-3-27B & 0.0158 & 0.0101 & \underline{0.0249} & \underline{0.0130} & 0.0408 & 0.0170 & 0.0767 & \underline{0.0242} \\
& Qwen2.5-VL-7B & 0.0159 & \underline{0.0101} & 0.0242 & 0.0128 & \textbf{0.0416} & \textbf{0.0172} & \textbf{0.0772} & \textbf{0.0242} \\
& Qwen2.5-VL-32B & 0.0159 & 0.0099 & 0.0246 & 0.0127 & \underline{0.0414} & 0.0169 & \underline{0.0768} & 0.0240 \\
& Qwen2.5-VL-72B & \underline{0.0161} & 0.0101 & \textbf{0.0252} & \textbf{0.0130} & 0.0406 & 0.0169 & 0.0766 & 0.0241 \\
\hline
& Ground Truth & 0.0157 & 0.0100 & 0.0248 & 0.0129 & 0.0418 & 0.0171 & 0.0770 & 0.0242 \\
\midrule
\multirow{7}{*}{Beauty} & Gemma-3-4B & 0.0081 & 0.0044 & 0.0178 & 0.0079 & 0.0264 & 0.0101 & 0.0507 & 0.0151 \\
& Gemma-3-12B & 0.0092 & 0.0051 & 0.0178 & 0.0081 & 0.0265 & 0.0103 & \underline{0.0519} & \underline{0.0155} \\
& Gemma-3-27B & \underline{0.0094} & \underline{0.0054} & 0.0175 & \underline{0.0083} & \underline{0.0281} & \underline{0.0109} & 0.0492 & 0.0152 \\
& Qwen2.5-VL-7B & 0.0081 & 0.0045 & \underline{0.0178} & 0.0079 & 0.0273 & 0.0103 & 0.0507 & 0.0151 \\
& Qwen2.5-VL-32B & \textbf{0.0119} & \textbf{0.0065} & \textbf{0.0203} & \textbf{0.0094} & \textbf{0.0337} & \textbf{0.0129} & \textbf{0.0595} & \textbf{0.0181} \\
& Qwen2.5-VL-72B & 0.0081 & 0.0044 & 0.0178 & 0.0078 & 0.0247 & 0.0097 & 0.0516 & 0.0152 \\
\hline
& Ground Truth & 0.0076 & 0.0045 & 0.0178 & 0.0080 & 0.0247 & 0.0098 & 0.0510 & 0.0153 \\
\midrule
\multirow{7}{*}{Electronics} & Gemma-3-4B & 0.0200 & \underline{0.0129} & 0.0310 & 0.0166 & 0.0547 & \textbf{0.0227} & 0.1044 & 0.0330 \\
& Gemma-3-12B & 0.0195 & 0.0127 & 0.0315 & 0.0167 & 0.0543 & 0.0225 & 0.1047 & 0.0329 \\
& Gemma-3-27B & 0.0193 & 0.0126 & 0.0315 & 0.0166 & \textbf{0.0549} & 0.0227 & 0.1045 & 0.0329 \\
& Qwen2.5-VL-7B & \underline{0.0200} & 0.0128 & 0.0311 & 0.0165 & 0.0547 & 0.0226 & \underline{0.1048} & \underline{0.0330} \\
& Qwen2.5-VL-32B & 0.0199 & 0.0128 & \underline{0.0321} & \underline{0.0168} & \underline{0.0549} & \underline{0.0227} & 0.1047 & 0.0330 \\
& Qwen2.5-VL-72B & \textbf{0.0210} & \textbf{0.0134} & \textbf{0.0332} & \textbf{0.0175} & 0.0532 & 0.0226 & \textbf{0.1067} & \textbf{0.0337} \\
\hline
& Ground Truth & 0.0203 & 0.0136 & 0.0329 & 0.0179 & 0.0551 & 0.0236 & 0.1088 & 0.0346 \\
\midrule
\multirow{7}{*}{Home} & Gemma-3-4B & 0.0104 & 0.0067 & 0.0169 & 0.0089 & 0.0281 & 0.0117 & 0.0558 & 0.0172 \\
& Gemma-3-12B & 0.0101 & \textbf{0.0072} & 0.0165 & \textbf{0.0093} & \textbf{0.0302} & \textbf{0.0128} & \textbf{0.0616} & \textbf{0.0191} \\
& Gemma-3-27B & \underline{0.0108} & \underline{0.0071} & \underline{0.0173} & \underline{0.0092} & \underline{0.0297} & \underline{0.0123} & \underline{0.0600} & \underline{0.0183} \\
& Qwen2.5-VL-7B & \textbf{0.0108} & 0.0069 & 0.0169 & 0.0090 & 0.0289 & 0.0120 & 0.0556 & 0.0173 \\
& Qwen2.5-VL-32B & 0.0102 & 0.0066 & 0.0171 & 0.0089 & 0.0276 & 0.0115 & 0.0580 & 0.0176 \\
& Qwen2.5-VL-72B & 0.0098 & 0.0063 & \textbf{0.0174} & 0.0088 & 0.0284 & 0.0116 & 0.0553 & 0.0170 \\
\hline
& Ground Truth & 0.0109 & 0.0069 & 0.0169 & 0.0088 & 0.0277 & 0.0116 & 0.0592 & 0.0179 \\
\midrule
\multirow{7}{*}{Instruments} & Gemma-3-4B & 0.0205 & 0.0139 & 0.0319 & 0.0177 & \underline{0.0505} & 0.0224 & 0.0872 & 0.0297 \\
& Gemma-3-12B & 0.0207 & \textbf{0.0142} & \textbf{0.0324} & \textbf{0.0181} & 0.0496 & 0.0224 & 0.0881 & 0.0301 \\
& Gemma-3-27B & 0.0201 & 0.0139 & 0.0314 & 0.0176 & 0.0501 & 0.0223 & 0.0875 & 0.0298 \\
& Qwen2.5-VL-7B & \underline{0.0207} & 0.0141 & 0.0314 & 0.0177 & 0.0499 & \underline{0.0224} & \underline{0.0889} & \textbf{0.0301} \\
& Qwen2.5-VL-32B & \textbf{0.0215} & \underline{0.0142} & \underline{0.0323} & \underline{0.0178} & \textbf{0.0510} & \textbf{0.0225} & \textbf{0.0893} & \underline{0.0301} \\
& Qwen2.5-VL-72B & 0.0205 & 0.0140 & 0.0315 & 0.0176 & 0.0501 & 0.0223 & 0.0887 & 0.0300 \\
\hline
& Ground Truth & 0.0219 & 0.0146 & 0.0319 & 0.0179 & 0.0519 & 0.0229 & 0.0913 & 0.0308 \\
\midrule
\multirow{7}{*}{Office} & Gemma-3-4B & 0.0183 & 0.0123 & 0.0295 & 0.0159 & 0.0453 & 0.0199 & 0.0822 & 0.0272 \\
& Gemma-3-12B & \textbf{0.0193} & \textbf{0.0127} & 0.0297 & 0.0161 & \textbf{0.0474} & \textbf{0.0205} & \textbf{0.0842} & \textbf{0.0278} \\
& Gemma-3-27B & 0.0180 & 0.0122 & 0.0292 & 0.0159 & 0.0458 & 0.0201 & 0.0822 & 0.0273 \\
& Qwen2.5-VL-7B & 0.0179 & 0.0122 & \textbf{0.0300} & \textbf{0.0161} & 0.0454 & 0.0199 & 0.0818 & 0.0272 \\
& Qwen2.5-VL-32B & \underline{0.0189} & \underline{0.0126} & 0.0297 & \underline{0.0161} & \underline{0.0474} & \underline{0.0205} & \underline{0.0836} & \underline{0.0277} \\
& Qwen2.5-VL-72B & 0.0184 & 0.0122 & \underline{0.0298} & 0.0159 & 0.0460 & 0.0200 & 0.0827 & 0.0272 \\
\hline
& Ground Truth & 0.0187 & 0.0122 & 0.0300 & 0.0159 & 0.0458 & 0.0199 & 0.0819 & 0.0270 \\
\midrule
\multirow{7}{*}{Scientific} & Gemma-3-4B & 0.0206 & 0.0134 & \underline{0.0328} & \underline{0.0173} & 0.0481 & 0.0212 & 0.0817 & \underline{0.0278} \\
& Gemma-3-12B & \underline{0.0211} & \textbf{0.0136} & 0.0315 & 0.0170 & 0.0480 & 0.0212 & 0.0806 & 0.0276 \\
& Gemma-3-27B & 0.0206 & 0.0134 & 0.0327 & 0.0173 & \textbf{0.0490} & \textbf{0.0214} & 0.0806 & 0.0276 \\
& Qwen2.5-VL-7B & \textbf{0.0212} & \underline{0.0135} & \textbf{0.0332} & \textbf{0.0174} & \underline{0.0486} & \underline{0.0213} & \textbf{0.0819} & \textbf{0.0278} \\
& Qwen2.5-VL-32B & 0.0204 & 0.0135 & 0.0319 & 0.0172 & 0.0474 & 0.0211 & 0.0793 & 0.0275 \\
& Qwen2.5-VL-72B & 0.0202 & 0.0132 & 0.0326 & 0.0172 & 0.0479 & 0.0211 & \underline{0.0817} & 0.0277 \\
\hline
& Ground Truth & 0.0214 & 0.0137 & 0.0323 & 0.0172 & 0.0478 & 0.0211 & 0.0804 & 0.0275 \\
\midrule
\multirow{7}{*}{Toys \& Games} & Gemma-3-4B & 0.0125 & 0.0080 & 0.0198 & 0.0103 & 0.0302 & 0.0129 & \underline{0.0564} & 0.0181 \\
& Gemma-3-12B & 0.0122 & 0.0079 & 0.0202 & 0.0104 & 0.0305 & 0.0130 & \textbf{0.0566} & 0.0181 \\
& Gemma-3-27B & 0.0125 & \underline{0.0083} & 0.0203 & \underline{0.0108} & 0.0324 & 0.0139 & 0.0558 & \textbf{0.0185} \\
& Qwen2.5-VL-7B & \textbf{0.0129} & \textbf{0.0083} & \underline{0.0205} & 0.0107 & \textbf{0.0331} & \textbf{0.0139} & 0.0558 & 0.0183 \\
& Qwen2.5-VL-32B & 0.0125 & 0.0080 & 0.0200 & 0.0103 & 0.0308 & 0.0131 & 0.0562 & 0.0181 \\
& Qwen2.5-VL-72B & \underline{0.0128} & 0.0082 & \textbf{0.0209} & \textbf{0.0108} & \underline{0.0330} & \underline{0.0139} & 0.0561 & \underline{0.0184} \\
\hline
& Ground Truth & 0.0124 & 0.0080 & 0.0201 & 0.0104 & 0.0307 & 0.0131 & 0.0565 & 0.0181 \\
\midrule
\multirow{7}{*}{Video Games} & Gemma-3-4B & 0.0237 & 0.0152 & 0.0398 & 0.0204 & 0.0647 & 0.0268 & 0.1186 & 0.0375 \\
& Gemma-3-12B & 0.0237 & 0.0151 & 0.0399 & 0.0204 & 0.0647 & 0.0267 & \textbf{0.1198} & 0.0377 \\
& Gemma-3-27B & \textbf{0.0246} & \textbf{0.0157} & 0.0392 & 0.0204 & 0.0645 & 0.0269 & 0.1191 & \underline{0.0378} \\
& Qwen2.5-VL-7B & 0.0237 & 0.0152 & \textbf{0.0409} & \textbf{0.0208} & \textbf{0.0649} & \underline{0.0269} & 0.1189 & 0.0377 \\
& Qwen2.5-VL-32B & 0.0235 & 0.0151 & \underline{0.0399} & \underline{0.0205} & \underline{0.0649} & 0.0269 & 0.1185 & 0.0375 \\
& Qwen2.5-VL-72B & \underline{0.0240} & \underline{0.0156} & 0.0387 & 0.0204 & 0.0645 & \textbf{0.0270} & \underline{0.1193} & \textbf{0.0379} \\
\hline
& Ground Truth & 0.0244 & 0.0158 & 0.0402 & 0.0209 & 0.0651 & 0.0273 & 0.1206 & 0.0384 \\
\bottomrule
\end{tabular}
\label{tab:performance_vbpr_i_to_t}
\end{table*}

\begin{table*}
\centering
\scriptsize
\caption{Performance Comparison for Model VBPR on Task Text-to-Image. R and N represent Recall and NDCG.}
\renewcommand{\arraystretch}{0.6}
\setlength{\tabcolsep}{5pt}
\begin{tabular}{@{}lc|cccccccc@{}}
\toprule
Dataset & Completion Method & R@5 & N@5 & R@10 & N@10 & R@20 & N@20 & R@50 & N@50 \\
\midrule
\multirow{7}{*}{Arts} & Gemma-3-4B & 0.0144 & 0.0093 & 0.0250 & 0.0127 & \textbf{0.0416} & \underline{0.0169} & 0.0738 & 0.0232 \\
& Gemma-3-12B & \textbf{0.0161} & \underline{0.0096} & 0.0251 & 0.0125 & \underline{0.0410} & 0.0165 & \textbf{0.0767} & 0.0236 \\
& Gemma-3-27B & 0.0095 & 0.0060 & 0.0174 & 0.0086 & 0.0297 & 0.0117 & 0.0641 & 0.0184 \\
& Qwen2.5-VL-7B & 0.0131 & 0.0080 & 0.0248 & 0.0118 & 0.0397 & 0.0155 & 0.0750 & 0.0225 \\
& Qwen2.5-VL-32B & 0.0146 & \textbf{0.0097} & \textbf{0.0262} & \textbf{0.0135} & 0.0408 & \textbf{0.0171} & 0.0751 & \textbf{0.0239} \\
& Qwen2.5-VL-72B & \underline{0.0151} & 0.0095 & \underline{0.0255} & \underline{0.0129} & 0.0403 & 0.0166 & \underline{0.0758} & \underline{0.0236} \\
\hline
& Ground Truth & 0.0157 & 0.0100 & 0.0248 & 0.0129 & 0.0418 & 0.0171 & 0.0770 & 0.0242 \\
\midrule
\multirow{7}{*}{Beauty} & Gemma-3-4B & 0.0072 & 0.0038 & 0.0109 & 0.0050 & 0.0195 & 0.0073 & 0.0488 & 0.0134 \\
& Gemma-3-12B & 0.0041 & 0.0023 & 0.0081 & 0.0036 & 0.0173 & 0.0060 & 0.0425 & 0.0110 \\
& Gemma-3-27B & \underline{0.0080} & \underline{0.0050} & 0.0128 & \underline{0.0066} & 0.0208 & \underline{0.0088} & 0.0506 & 0.0149 \\
& Qwen2.5-VL-7B & 0.0066 & 0.0036 & \textbf{0.0142} & 0.0061 & \underline{0.0248} & 0.0087 & \underline{0.0557} & \underline{0.0150} \\
& Qwen2.5-VL-32B & 0.0061 & 0.0037 & 0.0101 & 0.0051 & \textbf{0.0283} & \textbf{0.0098} & 0.0513 & 0.0145 \\
& Qwen2.5-VL-72B & \textbf{0.0089} & \textbf{0.0054} & \underline{0.0133} & \textbf{0.0069} & 0.0198 & 0.0085 & \textbf{0.0561} & \textbf{0.0160} \\
\hline
& Ground Truth & 0.0076 & 0.0045 & 0.0178 & 0.0080 & 0.0247 & 0.0098 & 0.0510 & 0.0153 \\
\midrule
\multirow{7}{*}{Electronics} & Gemma-3-4B & \textbf{0.0203} & \textbf{0.0135} & \textbf{0.0346} & \textbf{0.0182} & \textbf{0.0597} & \textbf{0.0248} & \underline{0.1090} & \textbf{0.0349} \\
& Gemma-3-12B & 0.0185 & 0.0121 & \underline{0.0328} & \underline{0.0168} & 0.0563 & \underline{0.0230} & 0.1048 & \underline{0.0331} \\
& Gemma-3-27B & \underline{0.0196} & \underline{0.0129} & 0.0299 & 0.0163 & 0.0509 & 0.0218 & 0.0998 & 0.0317 \\
& Qwen2.5-VL-7B & 0.0171 & 0.0107 & 0.0324 & 0.0159 & \underline{0.0574} & 0.0224 & \textbf{0.1091} & 0.0330 \\
& Qwen2.5-VL-32B & 0.0176 & 0.0111 & 0.0303 & 0.0155 & 0.0546 & 0.0218 & 0.1049 & 0.0322 \\
& Qwen2.5-VL-72B & 0.0165 & 0.0100 & 0.0296 & 0.0144 & 0.0518 & 0.0202 & 0.1036 & 0.0308 \\
\hline
& Ground Truth & 0.0203 & 0.0136 & 0.0329 & 0.0179 & 0.0551 & 0.0236 & 0.1088 & 0.0346 \\
\midrule
\multirow{7}{*}{Home} & Gemma-3-4B & 0.0100 & 0.0061 & 0.0177 & 0.0086 & 0.0310 & 0.0119 & 0.0622 & 0.0182 \\
& Gemma-3-12B & 0.0096 & 0.0060 & 0.0161 & 0.0081 & 0.0265 & 0.0108 & 0.0530 & 0.0160 \\
& Gemma-3-27B & \textbf{0.0123} & \underline{0.0078} & \textbf{0.0190} & \underline{0.0100} & \textbf{0.0327} & \underline{0.0135} & \textbf{0.0675} & \textbf{0.0205} \\
& Qwen2.5-VL-7B & 0.0110 & \textbf{0.0079} & \underline{0.0182} & \textbf{0.0103} & 0.0317 & \textbf{0.0137} & 0.0603 & \underline{0.0194} \\
& Qwen2.5-VL-32B & \underline{0.0112} & 0.0071 & 0.0181 & 0.0094 & \underline{0.0318} & 0.0129 & \underline{0.0646} & 0.0194 \\
& Qwen2.5-VL-72B & 0.0093 & 0.0062 & 0.0178 & 0.0089 & 0.0289 & 0.0117 & 0.0579 & 0.0174 \\
\hline
& Ground Truth & 0.0109 & 0.0069 & 0.0169 & 0.0088 & 0.0277 & 0.0116 & 0.0592 & 0.0179 \\
\midrule
\multirow{7}{*}{Instruments} & Gemma-3-4B & \underline{0.0206} & \underline{0.0135} & 0.0333 & \underline{0.0176} & 0.0524 & \underline{0.0224} & \underline{0.0939} & \underline{0.0307} \\
& Gemma-3-12B & 0.0199 & 0.0126 & 0.0324 & 0.0167 & 0.0520 & 0.0217 & 0.0922 & 0.0297 \\
& Gemma-3-27B & 0.0199 & 0.0131 & 0.0324 & 0.0171 & \textbf{0.0527} & 0.0222 & 0.0939 & 0.0304 \\
& Qwen2.5-VL-7B & 0.0189 & 0.0123 & 0.0307 & 0.0162 & 0.0498 & 0.0210 & 0.0916 & 0.0293 \\
& Qwen2.5-VL-32B & \textbf{0.0208} & \textbf{0.0141} & \underline{0.0336} & \textbf{0.0182} & 0.0513 & \textbf{0.0228} & 0.0923 & \textbf{0.0309} \\
& Qwen2.5-VL-72B & 0.0197 & 0.0127 & \textbf{0.0337} & 0.0172 & \underline{0.0524} & 0.0220 & \textbf{0.0940} & 0.0303 \\
\hline
& Ground Truth & 0.0219 & 0.0146 & 0.0319 & 0.0179 & 0.0519 & 0.0229 & 0.0913 & 0.0308 \\
\midrule
\multirow{7}{*}{Office} & Gemma-3-4B & 0.0148 & \underline{0.0097} & 0.0244 & \underline{0.0127} & 0.0382 & \underline{0.0162} & 0.0725 & \underline{0.0229} \\
& Gemma-3-12B & 0.0146 & 0.0093 & 0.0248 & 0.0126 & 0.0384 & 0.0160 & 0.0697 & 0.0222 \\
& Gemma-3-27B & \underline{0.0150} & 0.0095 & \underline{0.0251} & 0.0127 & 0.0389 & 0.0162 & 0.0715 & 0.0227 \\
& Qwen2.5-VL-7B & \textbf{0.0168} & \textbf{0.0112} & \textbf{0.0267} & \textbf{0.0144} & \textbf{0.0434} & \textbf{0.0185} & \textbf{0.0770} & \textbf{0.0252} \\
& Qwen2.5-VL-32B & 0.0124 & 0.0078 & 0.0228 & 0.0111 & 0.0389 & 0.0152 & 0.0736 & 0.0220 \\
& Qwen2.5-VL-72B & 0.0144 & 0.0089 & 0.0244 & 0.0121 & \underline{0.0395} & 0.0159 & \underline{0.0743} & 0.0228 \\
\hline
& Ground Truth & 0.0187 & 0.0122 & 0.0300 & 0.0159 & 0.0458 & 0.0199 & 0.0819 & 0.0270 \\
\midrule
\multirow{7}{*}{Scientific} & Gemma-3-4B & 0.0181 & \underline{0.0113} & 0.0289 & \underline{0.0148} & 0.0455 & 0.0190 & \textbf{0.0880} & \textbf{0.0273} \\
& Gemma-3-12B & \textbf{0.0192} & \textbf{0.0126} & \textbf{0.0302} & \textbf{0.0161} & 0.0445 & \underline{0.0197} & 0.0817 & \underline{0.0271} \\
& Gemma-3-27B & 0.0129 & 0.0080 & 0.0226 & 0.0111 & 0.0386 & 0.0151 & 0.0776 & 0.0228 \\
& Qwen2.5-VL-7B & 0.0140 & 0.0088 & 0.0240 & 0.0120 & 0.0379 & 0.0155 & 0.0764 & 0.0230 \\
& Qwen2.5-VL-32B & \underline{0.0183} & 0.0111 & 0.0272 & 0.0140 & \underline{0.0471} & 0.0190 & \underline{0.0834} & 0.0262 \\
& Qwen2.5-VL-72B & 0.0169 & 0.0106 & \underline{0.0292} & 0.0146 & \textbf{0.0496} & \textbf{0.0197} & 0.0832 & 0.0263 \\
\hline
& Ground Truth & 0.0214 & 0.0137 & 0.0323 & 0.0172 & 0.0478 & 0.0211 & 0.0804 & 0.0275 \\
\midrule
\multirow{7}{*}{Toys \& Games} & Gemma-3-4B & 0.0103 & 0.0063 & 0.0166 & 0.0083 & \underline{0.0288} & 0.0114 & 0.0524 & 0.0161 \\
& Gemma-3-12B & 0.0091 & 0.0060 & 0.0155 & 0.0081 & 0.0267 & 0.0109 & 0.0483 & 0.0152 \\
& Gemma-3-27B & \textbf{0.0118} & \textbf{0.0079} & \textbf{0.0185} & \textbf{0.0101} & 0.0274 & \textbf{0.0123} & 0.0521 & \underline{0.0172} \\
& Qwen2.5-VL-7B & 0.0094 & 0.0064 & 0.0152 & 0.0082 & 0.0264 & 0.0110 & 0.0474 & 0.0152 \\
& Qwen2.5-VL-32B & \underline{0.0107} & 0.0065 & 0.0181 & 0.0089 & 0.0281 & 0.0114 & \underline{0.0528} & 0.0162 \\
& Qwen2.5-VL-72B & 0.0105 & \underline{0.0070} & \underline{0.0182} & \underline{0.0095} & \textbf{0.0293} & \underline{0.0122} & \textbf{0.0546} & \textbf{0.0172} \\
\hline
& Ground Truth & 0.0124 & 0.0080 & 0.0201 & 0.0104 & 0.0307 & 0.0131 & 0.0565 & 0.0181 \\
\midrule
\multirow{7}{*}{Video Games} & Gemma-3-4B & 0.0236 & 0.0154 & 0.0396 & 0.0206 & 0.0645 & 0.0270 & 0.1193 & 0.0379 \\
& Gemma-3-12B & \textbf{0.0259} & 0.0164 & 0.0410 & 0.0214 & 0.0663 & \underline{0.0278} & 0.1196 & 0.0385 \\
& Gemma-3-27B & \underline{0.0257} & \underline{0.0168} & \underline{0.0413} & \underline{0.0219} & 0.0647 & 0.0278 & 0.1208 & \underline{0.0390} \\
& Qwen2.5-VL-7B & 0.0241 & 0.0155 & 0.0405 & 0.0208 & 0.0644 & 0.0269 & 0.1155 & 0.0371 \\
& Qwen2.5-VL-32B & 0.0242 & 0.0152 & 0.0404 & 0.0205 & \underline{0.0665} & 0.0272 & \underline{0.1225} & 0.0384 \\
& Qwen2.5-VL-72B & 0.0255 & \textbf{0.0169} & \textbf{0.0420} & \textbf{0.0222} & \textbf{0.0678} & \textbf{0.0288} & \textbf{0.1230} & \textbf{0.0398} \\
\hline
& Ground Truth & 0.0244 & 0.0158 & 0.0402 & 0.0209 & 0.0651 & 0.0273 & 0.1206 & 0.0384 \\
\bottomrule
\end{tabular}

\label{tab:performance_vbpr_t_to_i}
\end{table*}

\begin{table*}[htbp]
\centering
\scriptsize
\renewcommand{\arraystretch}{0.6}
\setlength{\tabcolsep}{5pt}
\caption{Performance Comparison for Model BM3 on Task Image-to-Text. R and N represent Recall and NDCG.}
\begin{tabular}{@{}lc|cccccccc@{}}
\toprule
Dataset & Completion Method & R@5 & N@5 & R@10 & N@10 & R@20 & N@20 & R@50 & N@50 \\
\midrule
\multirow{7}{*}{Arts} & Gemma-3-4B & \textbf{0.0560} & \underline{0.0344} & 0.0889 & \underline{0.0449} & 0.1371 & 0.0571 & 0.2239 & 0.0743 \\
& Gemma-3-12B & 0.0556 & 0.0342 & 0.0874 & 0.0445 & 0.1376 & 0.0572 & 0.2229 & 0.0741 \\
& Gemma-3-27B & 0.0554 & 0.0342 & 0.0873 & 0.0445 & 0.1374 & 0.0571 & 0.2253 & 0.0746 \\
& Qwen2.5-VL-7B & 0.0549 & 0.0338 & \underline{0.0889} & 0.0448 & \underline{0.1387} & \underline{0.0574} & 0.2246 & 0.0744 \\
& Qwen2.5-VL-32B & \underline{0.0558} & \textbf{0.0346} & 0.0881 & \textbf{0.0450} & 0.1386 & \textbf{0.0577} & \underline{0.2264} & \textbf{0.0751} \\
& Qwen2.5-VL-72B & 0.0557 & 0.0340 & \textbf{0.0890} & 0.0447 & \textbf{0.1391} & 0.0574 & \textbf{0.2265} & \underline{0.0746} \\
\hline
& Ground Truth & 0.0539 & 0.0334 & 0.0879 & 0.0443 & 0.1366 & 0.0566 & 0.2224 & 0.0737 \\
\midrule
\multirow{7}{*}{Beauty} & Gemma-3-4B & 0.0108 & 0.0070 & \textbf{0.0173} & \textbf{0.0091} & \textbf{0.0241} & \textbf{0.0109} & \textbf{0.0460} & \textbf{0.0152} \\
& Gemma-3-12B & 0.0112 & \textbf{0.0077} & 0.0145 & \underline{0.0088} & 0.0197 & \underline{0.0101} & 0.0376 & \underline{0.0137} \\
& Gemma-3-27B & 0.0112 & 0.0068 & 0.0126 & 0.0074 & 0.0211 & 0.0095 & \underline{0.0378} & 0.0129 \\
& Qwen2.5-VL-7B & \textbf{0.0112} & 0.0065 & \underline{0.0145} & 0.0075 & \underline{0.0220} & 0.0094 & 0.0340 & 0.0119 \\
& Qwen2.5-VL-32B & \underline{0.0112} & \underline{0.0077} & 0.0126 & 0.0082 & 0.0198 & 0.0100 & 0.0359 & 0.0134 \\
& Qwen2.5-VL-72B & 0.0103 & 0.0064 & 0.0126 & 0.0072 & 0.0197 & 0.0091 & 0.0359 & 0.0123 \\
\hline
& Ground Truth & 0.0104 & 0.0074 & 0.0123 & 0.0079 & 0.0198 & 0.0098 & 0.0363 & 0.0132 \\
\midrule
\multirow{7}{*}{Electronics} & Gemma-3-4B & 0.0236 & 0.0142 & \textbf{0.0406} & \underline{0.0199} & 0.0711 & 0.0278 & \textbf{0.1324} & \textbf{0.0403} \\
& Gemma-3-12B & \underline{0.0239} & 0.0143 & 0.0404 & 0.0198 & \textbf{0.0717} & \textbf{0.0280} & 0.1311 & 0.0401 \\
& Gemma-3-27B & \textbf{0.0239} & 0.0143 & 0.0403 & 0.0198 & \underline{0.0713} & \underline{0.0279} & \underline{0.1316} & 0.0401 \\
& Qwen2.5-VL-7B & 0.0238 & \textbf{0.0143} & \underline{0.0404} & \textbf{0.0199} & 0.0701 & 0.0276 & 0.1312 & \underline{0.0401} \\
& Qwen2.5-VL-32B & 0.0238 & 0.0143 & 0.0397 & 0.0196 & 0.0709 & 0.0277 & 0.1301 & 0.0398 \\
& Qwen2.5-VL-72B & 0.0237 & \underline{0.0143} & 0.0399 & 0.0197 & 0.0703 & 0.0277 & 0.1314 & 0.0401 \\
\hline
& Ground Truth & 0.0238 & 0.0145 & 0.0420 & 0.0206 & 0.0724 & 0.0285 & 0.1328 & 0.0408 \\
\midrule
\multirow{7}{*}{Home} & Gemma-3-4B & 0.0294 & 0.0185 & \underline{0.0521} & 0.0258 & 0.0818 & 0.0332 & 0.1531 & 0.0474 \\
& Gemma-3-12B & \underline{0.0296} & \underline{0.0186} & \textbf{0.0523} & \underline{0.0259} & 0.0818 & 0.0332 & \underline{0.1536} & \underline{0.0475} \\
& Gemma-3-27B & 0.0293 & 0.0185 & 0.0507 & 0.0254 & 0.0820 & 0.0332 & 0.1528 & 0.0473 \\
& Qwen2.5-VL-7B & \textbf{0.0303} & \textbf{0.0198} & 0.0504 & \textbf{0.0262} & \textbf{0.0825} & \textbf{0.0344} & \textbf{0.1540} & \textbf{0.0485} \\
& Qwen2.5-VL-32B & 0.0294 & 0.0185 & 0.0512 & 0.0255 & 0.0823 & 0.0334 & 0.1533 & 0.0474 \\
& Qwen2.5-VL-72B & 0.0293 & 0.0185 & 0.0507 & 0.0254 & \underline{0.0825} & \underline{0.0334} & 0.1528 & 0.0473 \\
\hline
& Ground Truth & 0.0293 & 0.0185 & 0.0521 & 0.0258 & 0.0815 & 0.0332 & 0.1528 & 0.0473 \\
\midrule
\multirow{7}{*}{Instruments} & Gemma-3-4B & \underline{0.0624} & 0.0398 & 0.0941 & 0.0500 & 0.1428 & 0.0623 & 0.2411 & 0.0818 \\
& Gemma-3-12B & 0.0623 & \underline{0.0408} & \textbf{0.0953} & \underline{0.0514} & \textbf{0.1437} & \underline{0.0637} & 0.2432 & \underline{0.0835} \\
& Gemma-3-27B & 0.0621 & 0.0401 & 0.0949 & 0.0507 & 0.1420 & 0.0627 & 0.2427 & 0.0827 \\
& Qwen2.5-VL-7B & 0.0621 & 0.0404 & \underline{0.0952} & 0.0510 & 0.1418 & 0.0629 & 0.2434 & 0.0830 \\
& Qwen2.5-VL-32B & 0.0618 & 0.0401 & 0.0947 & 0.0508 & 0.1428 & 0.0630 & \textbf{0.2454} & 0.0834 \\
& Qwen2.5-VL-72B & \textbf{0.0630} & \textbf{0.0412} & 0.0951 & \textbf{0.0515} & \underline{0.1431} & \textbf{0.0638} & \underline{0.2451} & \textbf{0.0840} \\
\hline
& Ground Truth & 0.0630 & 0.0408 & 0.0955 & 0.0512 & 0.1420 & 0.0631 & 0.2437 & 0.0833 \\
\midrule
\multirow{7}{*}{Office} & Gemma-3-4B & 0.0545 & 0.0348 & \underline{0.0853} & 0.0448 & \textbf{0.1284} & \underline{0.0556} & \underline{0.2054} & 0.0709 \\
& Gemma-3-12B & \underline{0.0554} & \textbf{0.0353} & \textbf{0.0854} & \textbf{0.0450} & \underline{0.1282} & \textbf{0.0557} & 0.2047 & 0.0709 \\
& Gemma-3-27B & 0.0549 & 0.0351 & 0.0846 & 0.0447 & 0.1277 & 0.0556 & 0.2045 & 0.0708 \\
& Qwen2.5-VL-7B & 0.0535 & 0.0341 & 0.0847 & 0.0441 & 0.1266 & 0.0547 & 0.2050 & 0.0702 \\
& Qwen2.5-VL-32B & \textbf{0.0555} & 0.0352 & 0.0852 & \underline{0.0448} & 0.1273 & 0.0555 & \textbf{0.2061} & \textbf{0.0711} \\
& Qwen2.5-VL-72B & 0.0552 & \underline{0.0352} & 0.0844 & 0.0447 & 0.1264 & 0.0553 & 0.2053 & \underline{0.0710} \\
\hline
& Ground Truth & 0.0544 & 0.0345 & 0.0848 & 0.0443 & 0.1278 & 0.0552 & 0.2048 & 0.0704 \\
\midrule
\multirow{7}{*}{Scientific} & Gemma-3-4B & 0.0434 & 0.0297 & \underline{0.0738} & \underline{0.0394} & \underline{0.1099} & 0.0485 & 0.1726 & 0.0609 \\
& Gemma-3-12B & \underline{0.0441} & \textbf{0.0303} & \textbf{0.0746} & \textbf{0.0400} & 0.1089 & \underline{0.0487} & 0.1734 & \textbf{0.0614} \\
& Gemma-3-27B & \textbf{0.0446} & \underline{0.0301} & 0.0737 & 0.0394 & 0.1083 & 0.0481 & 0.1722 & 0.0607 \\
& Qwen2.5-VL-7B & 0.0436 & 0.0298 & 0.0729 & 0.0392 & 0.1095 & 0.0484 & 0.1727 & 0.0609 \\
& Qwen2.5-VL-32B & 0.0438 & 0.0300 & 0.0727 & 0.0393 & \textbf{0.1102} & \textbf{0.0488} & \textbf{0.1741} & \underline{0.0614} \\
& Qwen2.5-VL-72B & 0.0434 & 0.0295 & 0.0725 & 0.0388 & 0.1096 & 0.0482 & \underline{0.1734} & 0.0608 \\
\hline
& Ground Truth & 0.0445 & 0.0304 & 0.0737 & 0.0398 & 0.1089 & 0.0487 & 0.1746 & 0.0616 \\
\midrule
\multirow{7}{*}{Toys \& Games} & Gemma-3-4B & 0.0342 & \underline{0.0232} & 0.0511 & \underline{0.0286} & \underline{0.0759} & \underline{0.0349} & \textbf{0.1408} & \textbf{0.0476} \\
& Gemma-3-12B & 0.0335 & 0.0225 & 0.0498 & 0.0278 & 0.0754 & 0.0342 & 0.1379 & 0.0465 \\
& Gemma-3-27B & \textbf{0.0348} & \textbf{0.0233} & 0.0513 & 0.0286 & \textbf{0.0767} & \textbf{0.0349} & \underline{0.1398} & 0.0473 \\
& Qwen2.5-VL-7B & 0.0339 & 0.0228 & \underline{0.0514} & 0.0284 & 0.0758 & 0.0345 & 0.1385 & 0.0469 \\
& Qwen2.5-VL-32B & \underline{0.0343} & 0.0231 & 0.0510 & 0.0285 & 0.0757 & 0.0348 & 0.1395 & \underline{0.0473} \\
& Qwen2.5-VL-72B & 0.0337 & 0.0228 & \textbf{0.0520} & \textbf{0.0287} & 0.0756 & 0.0346 & 0.1391 & 0.0471 \\
\hline
& Ground Truth & 0.0351 & 0.0233 & 0.0520 & 0.0287 & 0.0765 & 0.0349 & 0.1406 & 0.0476 \\
\midrule
\multirow{7}{*}{Video Games} & Gemma-3-4B & 0.0582 & 0.0384 & 0.0933 & 0.0497 & 0.1425 & 0.0622 & 0.2352 & 0.0807 \\
& Gemma-3-12B & 0.0578 & 0.0382 & 0.0940 & 0.0499 & \underline{0.1432} & 0.0624 & 0.2344 & 0.0806 \\
& Gemma-3-27B & 0.0586 & 0.0385 & 0.0938 & 0.0498 & 0.1427 & 0.0623 & 0.2347 & 0.0806 \\
& Qwen2.5-VL-7B & \textbf{0.0590} & \textbf{0.0391} & \textbf{0.0945} & \textbf{0.0505} & \textbf{0.1436} & \textbf{0.0630} & \textbf{0.2360} & \textbf{0.0814} \\
& Qwen2.5-VL-32B & 0.0580 & 0.0385 & \underline{0.0940} & \underline{0.0501} & 0.1421 & \underline{0.0624} & \underline{0.2354} & \underline{0.0809} \\
& Qwen2.5-VL-72B & \underline{0.0586} & \underline{0.0386} & 0.0939 & 0.0500 & 0.1422 & 0.0623 & 0.2345 & 0.0807 \\
\hline
& Ground Truth & 0.0594 & 0.0391 & 0.0944 & 0.0504 & 0.1434 & 0.0629 & 0.2362 & 0.0814 \\
\bottomrule
\end{tabular}

\label{tab:performance_bm3_i_to_t}
\end{table*}

\begin{table*}
\centering
\caption{Performance Comparison for Model BM3 on Task Text-to-Image. R and N represent Recall and NDCG.}
\scriptsize
\renewcommand{\arraystretch}{0.6}
\setlength{\tabcolsep}{5pt}
\begin{tabular}{@{}lc|cccccccc@{}}
\toprule
Dataset & Completion Method & R@5 & N@5 & R@10 & N@10 & R@20 & N@20 & R@50 & N@50 \\
\midrule
\multirow{7}{*}{Arts} & Gemma-3-4B & 0.0559 & 0.0348 & 0.0881 & 0.0452 & 0.1402 & 0.0584 & \textbf{0.2271} & \underline{0.0756} \\
& Gemma-3-12B & \underline{0.0562} & \textbf{0.0359} & 0.0903 & \textbf{0.0469} & 0.1403 & \textbf{0.0595} & 0.2226 & \textbf{0.0759} \\
& Gemma-3-27B & 0.0561 & 0.0350 & \underline{0.0908} & 0.0462 & 0.1395 & 0.0585 & 0.2205 & 0.0747 \\
& Qwen2.5-VL-7B & \textbf{0.0566} & \underline{0.0355} & 0.0902 & \underline{0.0463} & \underline{0.1403} & \underline{0.0590} & 0.2213 & 0.0752 \\
& Qwen2.5-VL-32B & 0.0560 & 0.0344 & \textbf{0.0920} & 0.0460 & \textbf{0.1405} & 0.0582 & \underline{0.2260} & 0.0752 \\
& Qwen2.5-VL-72B & 0.0549 & 0.0344 & 0.0893 & 0.0455 & 0.1392 & 0.0581 & 0.2239 & 0.0750 \\
\hline
& Ground Truth & 0.0539 & 0.0334 & 0.0879 & 0.0443 & 0.1366 & 0.0566 & 0.2224 & 0.0737 \\
\midrule
\multirow{7}{*}{Beauty} & Gemma-3-4B & 0.0103 & 0.0066 & 0.0131 & 0.0075 & \underline{0.0222} & 0.0098 & \textbf{0.0432} & \textbf{0.0140} \\
& Gemma-3-12B & \underline{0.0104} & \textbf{0.0075} & \underline{0.0151} & \textbf{0.0089} & 0.0202 & \textbf{0.0102} & 0.0345 & 0.0131 \\
& Gemma-3-27B & 0.0095 & \underline{0.0071} & 0.0141 & \underline{0.0084} & 0.0204 & \underline{0.0100} & 0.0349 & 0.0129 \\
& Qwen2.5-VL-7B & 0.0103 & 0.0063 & 0.0123 & 0.0069 & 0.0212 & 0.0092 & \underline{0.0422} & \underline{0.0135} \\
& Qwen2.5-VL-32B & \textbf{0.0122} & 0.0070 & 0.0146 & 0.0078 & \textbf{0.0230} & 0.0099 & 0.0404 & 0.0134 \\
& Qwen2.5-VL-72B & 0.0103 & 0.0061 & \textbf{0.0151} & 0.0077 & 0.0221 & 0.0094 & 0.0408 & 0.0131 \\
\hline
& Ground Truth & 0.0104 & 0.0074 & 0.0123 & 0.0079 & 0.0198 & 0.0098 & 0.0363 & 0.0132 \\
\midrule
\multirow{7}{*}{Electronics} & Gemma-3-4B & 0.0242 & 0.0148 & 0.0429 & \underline{0.0211} & 0.0727 & 0.0289 & 0.1334 & 0.0413 \\
& Gemma-3-12B & 0.0243 & 0.0148 & \textbf{0.0431} & 0.0211 & 0.0727 & 0.0289 & \underline{0.1341} & \underline{0.0414} \\
& Gemma-3-27B & 0.0243 & 0.0147 & 0.0430 & 0.0210 & 0.0716 & 0.0285 & 0.1332 & 0.0411 \\
& Qwen2.5-VL-7B & \textbf{0.0250} & \textbf{0.0151} & \underline{0.0430} & \textbf{0.0211} & 0.0727 & \textbf{0.0289} & \textbf{0.1344} & \textbf{0.0415} \\
& Qwen2.5-VL-32B & 0.0242 & 0.0147 & 0.0366 & 0.0188 & \underline{0.0730} & 0.0283 & 0.1290 & 0.0396 \\
& Qwen2.5-VL-72B & \underline{0.0243} & \underline{0.0148} & 0.0427 & 0.0210 & \textbf{0.0731} & \underline{0.0289} & 0.1340 & 0.0413 \\
\hline
& Ground Truth & 0.0238 & 0.0145 & 0.0420 & 0.0206 & 0.0724 & 0.0285 & 0.1328 & 0.0408 \\
\midrule
\multirow{7}{*}{Home} & Gemma-3-4B & \underline{0.0299} & \underline{0.0205} & 0.0489 & 0.0266 & \underline{0.0817} & \underline{0.0349} & 0.1507 & 0.0487 \\
& Gemma-3-12B & 0.0299 & \textbf{0.0210} & 0.0491 & \textbf{0.0272} & \textbf{0.0826} & \textbf{0.0357} & \textbf{0.1516} & \textbf{0.0494} \\
& Gemma-3-27B & 0.0299 & 0.0202 & 0.0484 & 0.0261 & 0.0812 & 0.0345 & 0.1503 & 0.0483 \\
& Qwen2.5-VL-7B & 0.0296 & 0.0203 & \underline{0.0491} & \underline{0.0266} & 0.0808 & 0.0347 & 0.1505 & 0.0486 \\
& Qwen2.5-VL-32B & 0.0296 & 0.0203 & 0.0490 & 0.0266 & 0.0812 & 0.0348 & \underline{0.1511} & \underline{0.0488} \\
& Qwen2.5-VL-72B & \textbf{0.0301} & 0.0202 & \textbf{0.0492} & 0.0263 & 0.0805 & 0.0343 & 0.1509 & 0.0484 \\
\hline
& Ground Truth & 0.0293 & 0.0185 & 0.0521 & 0.0258 & 0.0815 & 0.0332 & 0.1528 & 0.0473 \\
\midrule
\multirow{7}{*}{Instruments} & Gemma-3-4B & \textbf{0.0654} & \textbf{0.0420} & \textbf{0.0981} & \textbf{0.0525} & 0.1449 & \textbf{0.0644} & 0.2482 & \textbf{0.0849} \\
& Gemma-3-12B & 0.0639 & 0.0414 & 0.0967 & 0.0520 & 0.1446 & 0.0641 & 0.2466 & 0.0844 \\
& Gemma-3-27B & 0.0635 & 0.0413 & 0.0971 & \underline{0.0521} & \underline{0.1451} & \underline{0.0642} & 0.2479 & \underline{0.0846} \\
& Qwen2.5-VL-7B & 0.0631 & 0.0407 & 0.0972 & 0.0517 & 0.1447 & 0.0637 & \underline{0.2486} & 0.0844 \\
& Qwen2.5-VL-32B & \underline{0.0640} & \underline{0.0414} & 0.0949 & 0.0513 & 0.1449 & 0.0640 & 0.2464 & 0.0841 \\
& Qwen2.5-VL-72B & 0.0630 & 0.0405 & \underline{0.0980} & 0.0518 & \textbf{0.1458} & 0.0640 & \textbf{0.2487} & 0.0845 \\
\hline
& Ground Truth & 0.0630 & 0.0408 & 0.0955 & 0.0512 & 0.1420 & 0.0631 & 0.2437 & 0.0833 \\
\midrule
\multirow{7}{*}{Office} & Gemma-3-4B & 0.0549 & \underline{0.0349} & \textbf{0.0870} & \underline{0.0453} & 0.1284 & 0.0557 & 0.2041 & 0.0708 \\
& Gemma-3-12B & 0.0542 & 0.0343 & 0.0854 & 0.0444 & 0.1285 & 0.0553 & 0.2039 & 0.0703 \\
& Gemma-3-27B & 0.0539 & 0.0343 & \underline{0.0868} & 0.0449 & 0.1269 & 0.0551 & 0.2039 & 0.0704 \\
& Qwen2.5-VL-7B & 0.0547 & 0.0347 & 0.0861 & 0.0449 & \textbf{0.1304} & \underline{0.0561} & \textbf{0.2057} & \underline{0.0710} \\
& Qwen2.5-VL-32B & \textbf{0.0554} & \textbf{0.0354} & 0.0865 & \textbf{0.0455} & \underline{0.1298} & \textbf{0.0564} & \underline{0.2049} & \textbf{0.0712} \\
& Qwen2.5-VL-72B & \underline{0.0552} & 0.0347 & 0.0864 & 0.0448 & 0.1291 & 0.0555 & 0.2031 & 0.0702 \\
\hline
& Ground Truth & 0.0544 & 0.0345 & 0.0848 & 0.0443 & 0.1278 & 0.0552 & 0.2048 & 0.0704 \\
\midrule
\multirow{7}{*}{Scientific} & Gemma-3-4B & 0.0455 & 0.0303 & 0.0737 & 0.0393 & 0.1087 & 0.0483 & 0.1760 & 0.0615 \\
& Gemma-3-12B & \textbf{0.0476} & \textbf{0.0321} & 0.0704 & 0.0394 & 0.1041 & 0.0479 & 0.1697 & 0.0609 \\
& Gemma-3-27B & 0.0466 & 0.0309 & 0.0731 & 0.0395 & 0.1097 & 0.0487 & \textbf{0.1799} & 0.0625 \\
& Qwen2.5-VL-7B & \underline{0.0472} & \underline{0.0317} & \textbf{0.0752} & \textbf{0.0406} & \textbf{0.1116} & \underline{0.0498} & \underline{0.1769} & \textbf{0.0627} \\
& Qwen2.5-VL-32B & 0.0451 & 0.0303 & 0.0734 & 0.0394 & 0.1064 & 0.0478 & 0.1736 & 0.0610 \\
& Qwen2.5-VL-72B & 0.0463 & 0.0316 & \underline{0.0738} & \underline{0.0404} & \underline{0.1113} & \textbf{0.0499} & 0.1759 & \underline{0.0626} \\
\hline
& Ground Truth & 0.0445 & 0.0304 & 0.0737 & 0.0398 & 0.1089 & 0.0487 & 0.1746 & 0.0616 \\
\midrule
\multirow{7}{*}{Toys \& Games} & Gemma-3-4B & 0.0332 & 0.0223 & 0.0512 & 0.0281 & 0.0772 & 0.0346 & 0.1392 & 0.0468 \\
& Gemma-3-12B & \textbf{0.0340} & \textbf{0.0227} & \textbf{0.0523} & \textbf{0.0286} & 0.0775 & \textbf{0.0349} & 0.1391 & \textbf{0.0470} \\
& Gemma-3-27B & 0.0327 & 0.0218 & 0.0512 & 0.0277 & \textbf{0.0784} & 0.0345 & \underline{0.1397} & 0.0466 \\
& Qwen2.5-VL-7B & \underline{0.0339} & \underline{0.0224} & 0.0514 & 0.0280 & 0.0759 & 0.0342 & \textbf{0.1404} & \underline{0.0469} \\
& Qwen2.5-VL-32B & 0.0331 & 0.0224 & \underline{0.0516} & \underline{0.0284} & 0.0768 & 0.0347 & 0.1386 & 0.0469 \\
& Qwen2.5-VL-72B & 0.0334 & 0.0224 & 0.0507 & 0.0280 & \underline{0.0781} & \underline{0.0348} & 0.1390 & 0.0468 \\
\hline
& Ground Truth & 0.0351 & 0.0233 & 0.0520 & 0.0287 & 0.0765 & 0.0349 & 0.1406 & 0.0476 \\
\midrule
\multirow{7}{*}{Video Games} & Gemma-3-4B & \underline{0.0615} & \underline{0.0400} & 0.0970 & 0.0515 & 0.1462 & 0.0640 & 0.2369 & 0.0821 \\
& Gemma-3-12B & \textbf{0.0623} & \textbf{0.0400} & \underline{0.0981} & \underline{0.0516} & \underline{0.1474} & \underline{0.0641} & \underline{0.2408} & \underline{0.0828} \\
& Gemma-3-27B & 0.0615 & 0.0398 & \textbf{0.0982} & \textbf{0.0517} & \textbf{0.1475} & \textbf{0.0642} & \textbf{0.2430} & \textbf{0.0833} \\
& Qwen2.5-VL-7B & 0.0608 & 0.0399 & 0.0963 & 0.0513 & 0.1456 & 0.0639 & 0.2388 & 0.0825 \\
& Qwen2.5-VL-32B & 0.0602 & 0.0395 & 0.0972 & 0.0515 & 0.1450 & 0.0636 & 0.2370 & 0.0820 \\
& Qwen2.5-VL-72B & 0.0574 & 0.0377 & 0.0939 & 0.0495 & 0.1424 & 0.0618 & 0.2369 & 0.0807 \\
\hline
& Ground Truth & 0.0594 & 0.0391 & 0.0944 & 0.0504 & 0.1434 & 0.0629 & 0.2362 & 0.0814 \\
\bottomrule
\end{tabular}

\label{tab:performance_bm3_t_to_i}
\end{table*}

\begin{table*}
\centering
\caption{Performance Comparison for Model FREEDOM on Task Image-to-Text. R and N represent Recall and NDCG.}
\scriptsize
\renewcommand{\arraystretch}{0.6}
\setlength{\tabcolsep}{5pt}
\begin{tabular}{@{}lc|cccccccc@{}}
\toprule
Dataset & Completion Method & R@5 & N@5 & R@10 & N@10 & R@20 & N@20 & R@50 & N@50 \\
\midrule
\multirow{7}{*}{Arts} & Gemma-3-4B & 0.0501 & 0.0323 & 0.0880 & 0.0445 & \underline{0.1424} & 0.0583 & 0.2301 & 0.0757 \\
& Gemma-3-12B & \underline{0.0522} & 0.0332 & 0.0871 & 0.0445 & 0.1372 & 0.0570 & \underline{0.2316} & 0.0758 \\
& Gemma-3-27B & \textbf{0.0533} & \underline{0.0334} & \textbf{0.0893} & \underline{0.0450} & \textbf{0.1444} & \textbf{0.0589} & 0.2307 & \underline{0.0761} \\
& Qwen2.5-VL-7B & 0.0515 & 0.0332 & \underline{0.0886} & \textbf{0.0450} & 0.1370 & 0.0571 & 0.2287 & 0.0755 \\
& Qwen2.5-VL-32B & 0.0493 & 0.0316 & 0.0859 & 0.0434 & 0.1364 & 0.0562 & 0.2290 & 0.0746 \\
& Qwen2.5-VL-72B & 0.0521 & \textbf{0.0336} & 0.0845 & 0.0440 & 0.1416 & \underline{0.0585} & \textbf{0.2329} & \textbf{0.0766} \\
\hline
& Ground Truth & 0.0545 & 0.0349 & 0.0912 & 0.0467 & 0.1446 & 0.0603 & 0.2341 & 0.0782 \\
\midrule
\multirow{7}{*}{Beauty} & Gemma-3-4B & 0.0140 & 0.0100 & 0.0223 & 0.0126 & 0.0378 & 0.0167 & 0.0614 & 0.0213 \\
& Gemma-3-12B & 0.0159 & 0.0099 & 0.0265 & 0.0134 & 0.0358 & 0.0158 & \underline{0.0716} & 0.0232 \\
& Gemma-3-27B & 0.0176 & 0.0113 & 0.0265 & 0.0143 & 0.0409 & \underline{0.0179} & \textbf{0.0776} & \textbf{0.0255} \\
& Qwen2.5-VL-7B & \textbf{0.0197} & \textbf{0.0135} & \textbf{0.0331} & \textbf{0.0180} & \textbf{0.0421} & \textbf{0.0203} & 0.0575 & \underline{0.0236} \\
& Qwen2.5-VL-32B & \underline{0.0187} & \underline{0.0117} & 0.0256 & 0.0139 & 0.0366 & 0.0168 & 0.0609 & 0.0218 \\
& Qwen2.5-VL-72B & 0.0154 & 0.0101 & \underline{0.0300} & \underline{0.0147} & \underline{0.0417} & 0.0177 & 0.0652 & 0.0225 \\
\hline
& Ground Truth & 0.0187 & 0.0106 & 0.0314 & 0.0147 & 0.0487 & 0.0192 & 0.0844 & 0.0264 \\
\midrule
\multirow{7}{*}{Electronics} & Gemma-3-4B & \textbf{0.0319} & \textbf{0.0198} & 0.0456 & \textbf{0.0243} & 0.0792 & \textbf{0.0332} & 0.1533 & \underline{0.0483} \\
& Gemma-3-12B & 0.0269 & 0.0163 & \textbf{0.0475} & 0.0232 & 0.0790 & 0.0314 & \underline{0.1565} & 0.0472 \\
& Gemma-3-27B & 0.0238 & 0.0147 & 0.0389 & 0.0197 & 0.0729 & 0.0286 & 0.1403 & 0.0424 \\
& Qwen2.5-VL-7B & 0.0231 & 0.0150 & 0.0407 & 0.0207 & \underline{0.0802} & 0.0309 & \textbf{0.1579} & 0.0467 \\
& Qwen2.5-VL-32B & 0.0229 & 0.0149 & 0.0410 & 0.0209 & 0.0791 & 0.0309 & 0.1464 & 0.0447 \\
& Qwen2.5-VL-72B & \underline{0.0286} & \underline{0.0181} & \underline{0.0464} & \underline{0.0239} & \textbf{0.0810} & \underline{0.0330} & 0.1564 & \textbf{0.0486} \\
\hline
& Ground Truth & 0.0309 & 0.0191 & 0.0541 & 0.0267 & 0.0885 & 0.0358 & 0.1543 & 0.0492 \\
\midrule
\multirow{7}{*}{Home} & Gemma-3-4B & 0.0314 & 0.0214 & 0.0494 & 0.0273 & 0.0838 & 0.0360 & 0.1511 & 0.0495 \\
& Gemma-3-12B & 0.0320 & 0.0217 & 0.0505 & 0.0277 & 0.0823 & 0.0358 & \underline{0.1537} & 0.0501 \\
& Gemma-3-27B & 0.0323 & 0.0219 & 0.0495 & 0.0275 & \textbf{0.0839} & 0.0362 & 0.1528 & 0.0500 \\
& Qwen2.5-VL-7B & 0.0322 & 0.0219 & 0.0501 & 0.0276 & 0.0833 & 0.0361 & \textbf{0.1581} & \textbf{0.0510} \\
& Qwen2.5-VL-32B & \underline{0.0327} & \underline{0.0222} & \textbf{0.0523} & \textbf{0.0286} & \underline{0.0839} & \textbf{0.0366} & 0.1534 & 0.0505 \\
& Qwen2.5-VL-72B & \textbf{0.0327} & \textbf{0.0223} & \underline{0.0515} & \underline{0.0285} & 0.0833 & \underline{0.0365} & 0.1528 & \underline{0.0505} \\
\hline
& Ground Truth & 0.0312 & 0.0215 & 0.0500 & 0.0276 & 0.0803 & 0.0353 & 0.1544 & 0.0501 \\
\midrule
\multirow{7}{*}{Instruments} & Gemma-3-4B & 0.0651 & \textbf{0.0426} & \textbf{0.1024} & \textbf{0.0547} & 0.1503 & \textbf{0.0668} & \underline{0.2515} & \textbf{0.0870} \\
& Gemma-3-12B & \textbf{0.0656} & 0.0417 & 0.1016 & 0.0534 & 0.1502 & 0.0657 & 0.2504 & 0.0856 \\
& Gemma-3-27B & 0.0634 & 0.0410 & 0.1012 & 0.0531 & 0.1504 & 0.0656 & 0.2504 & 0.0854 \\
& Qwen2.5-VL-7B & 0.0635 & 0.0415 & 0.1009 & 0.0537 & \textbf{0.1515} & \underline{0.0665} & \textbf{0.2519} & \underline{0.0865} \\
& Qwen2.5-VL-32B & 0.0640 & 0.0416 & \underline{0.1020} & \underline{0.0539} & 0.1497 & 0.0660 & 0.2506 & 0.0861 \\
& Qwen2.5-VL-72B & \underline{0.0653} & \underline{0.0424} & 0.1004 & 0.0537 & \underline{0.1508} & 0.0665 & 0.2513 & 0.0865 \\
\hline
& Ground Truth & 0.0651 & 0.0425 & 0.1031 & 0.0548 & 0.1539 & 0.0676 & 0.2544 & 0.0877 \\
\midrule
\multirow{7}{*}{Office} & Gemma-3-4B & \underline{0.0509} & \underline{0.0328} & \underline{0.0818} & \textbf{0.0428} & \textbf{0.1251} & \textbf{0.0537} & 0.2007 & 0.0687 \\
& Gemma-3-12B & 0.0509 & 0.0325 & \textbf{0.0824} & \underline{0.0427} & 0.1245 & 0.0533 & \textbf{0.2044} & \textbf{0.0691} \\
& Gemma-3-27B & \textbf{0.0514} & \textbf{0.0328} & 0.0804 & 0.0422 & 0.1246 & \underline{0.0533} & \underline{0.2028} & \underline{0.0688} \\
& Qwen2.5-VL-7B & 0.0507 & 0.0319 & 0.0813 & 0.0417 & \underline{0.1250} & 0.0528 & 0.1997 & 0.0676 \\
& Qwen2.5-VL-32B & 0.0497 & 0.0319 & 0.0810 & 0.0420 & 0.1245 & 0.0530 & 0.2017 & 0.0683 \\
& Qwen2.5-VL-72B & 0.0497 & 0.0318 & 0.0800 & 0.0415 & 0.1233 & 0.0524 & 0.2009 & 0.0678 \\
\hline
& Ground Truth & 0.0533 & 0.0342 & 0.0859 & 0.0448 & 0.1279 & 0.0554 & 0.2065 & 0.0709 \\
\midrule
\multirow{7}{*}{Scientific} & Gemma-3-4B & \underline{0.0515} & 0.0342 & \underline{0.0798} & \underline{0.0434} & 0.1178 & \underline{0.0529} & 0.1796 & 0.0652 \\
& Gemma-3-12B & 0.0509 & 0.0333 & 0.0761 & 0.0415 & 0.1155 & 0.0514 & \underline{0.1844} & 0.0651 \\
& Gemma-3-27B & 0.0515 & \underline{0.0345} & 0.0793 & \textbf{0.0436} & 0.1158 & 0.0528 & 0.1840 & \textbf{0.0664} \\
& Qwen2.5-VL-7B & \textbf{0.0525} & \textbf{0.0347} & 0.0778 & 0.0429 & \underline{0.1198} & \textbf{0.0534} & 0.1811 & 0.0655 \\
& Qwen2.5-VL-32B & 0.0514 & 0.0332 & 0.0767 & 0.0414 & 0.1143 & 0.0507 & 0.1817 & 0.0641 \\
& Qwen2.5-VL-72B & 0.0512 & 0.0330 & \textbf{0.0799} & 0.0423 & \textbf{0.1204} & 0.0525 & \textbf{0.1864} & \underline{0.0656} \\
\hline
& Ground Truth & 0.0541 & 0.0348 & 0.0788 & 0.0428 & 0.1180 & 0.0527 & 0.1908 & 0.0672 \\
\midrule
\multirow{7}{*}{Toys \& Games} & Gemma-3-4B & 0.0355 & 0.0236 & 0.0534 & 0.0294 & 0.0797 & 0.0360 & 0.1356 & 0.0470 \\
& Gemma-3-12B & 0.0351 & 0.0235 & 0.0543 & 0.0297 & \underline{0.0817} & 0.0366 & \underline{0.1425} & \underline{0.0486} \\
& Gemma-3-27B & \textbf{0.0374} & \textbf{0.0251} & \textbf{0.0558} & \textbf{0.0310} & \textbf{0.0844} & \textbf{0.0382} & \textbf{0.1434} & \textbf{0.0499} \\
& Qwen2.5-VL-7B & \underline{0.0365} & \underline{0.0247} & 0.0532 & \underline{0.0301} & 0.0796 & \underline{0.0367} & 0.1383 & 0.0483 \\
& Qwen2.5-VL-32B & 0.0352 & 0.0234 & 0.0544 & 0.0296 & 0.0800 & 0.0360 & 0.1408 & 0.0479 \\
& Qwen2.5-VL-72B & 0.0364 & 0.0239 & \underline{0.0547} & 0.0298 & 0.0808 & 0.0364 & 0.1352 & 0.0471 \\
\hline
& Ground Truth & 0.0376 & 0.0255 & 0.0559 & 0.0314 & 0.0851 & 0.0388 & 0.1452 & 0.0506 \\
\midrule
\multirow{7}{*}{Video Games} & Gemma-3-4B & \textbf{0.0638} & \textbf{0.0417} & 0.0997 & \underline{0.0533} & 0.1528 & \underline{0.0668} & \textbf{0.2530} & \underline{0.0869} \\
& Gemma-3-12B & \underline{0.0625} & 0.0409 & 0.0996 & 0.0530 & 0.1524 & 0.0664 & 0.2505 & 0.0860 \\
& Gemma-3-27B & 0.0624 & \underline{0.0413} & 0.0995 & \textbf{0.0533} & \textbf{0.1550} & \textbf{0.0675} & \underline{0.2528} & \textbf{0.0870} \\
& Qwen2.5-VL-7B & 0.0615 & 0.0403 & \underline{0.1000} & 0.0528 & \underline{0.1549} & 0.0667 & 0.2496 & 0.0857 \\
& Qwen2.5-VL-32B & 0.0622 & 0.0403 & 0.0979 & 0.0518 & 0.1538 & 0.0660 & 0.2516 & 0.0856 \\
& Qwen2.5-VL-72B & 0.0614 & 0.0402 & \textbf{0.1005} & 0.0528 & 0.1526 & 0.0661 & 0.2514 & 0.0859 \\
\hline
& Ground Truth & 0.0623 & 0.0404 & 0.0985 & 0.0521 & 0.1523 & 0.0658 & 0.2508 & 0.0855 \\
\bottomrule
\end{tabular}

\label{tab:performance_freedom_i_to_t}
\end{table*}

\begin{table*}
\centering
\scriptsize
\renewcommand{\arraystretch}{0.6}
\setlength{\tabcolsep}{5pt}
\caption{Performance Comparison for Model FREEDOM on Task Text-to-Image. R and N represent Recall and NDCG.}
\begin{tabular}{@{}lc|cccccccc@{}}
\toprule
Dataset & Completion Method & R@5 & N@5 & R@10 & N@10 & R@20 & N@20 & R@50 & N@50 \\
\midrule
\multirow{7}{*}{Arts} & Gemma-3-4B & 0.0551 & 0.0346 & 0.0919 & 0.0464 & 0.1445 & 0.0597 & 0.2348 & 0.0776 \\
& Gemma-3-12B & \textbf{0.0561} & \underline{0.0349} & 0.0902 & 0.0459 & 0.1440 & 0.0595 & 0.2352 & 0.0777 \\
& Gemma-3-27B & 0.0547 & 0.0348 & \underline{0.0923} & \textbf{0.0470} & \textbf{0.1458} & \textbf{0.0605} & \textbf{0.2366} & \textbf{0.0785} \\
& Qwen2.5-VL-7B & 0.0547 & 0.0341 & 0.0916 & 0.0459 & 0.1450 & 0.0594 & \underline{0.2363} & 0.0776 \\
& Qwen2.5-VL-32B & 0.0554 & 0.0348 & \textbf{0.0926} & 0.0467 & 0.1445 & 0.0598 & 0.2352 & 0.0779 \\
& Qwen2.5-VL-72B & \underline{0.0554} & \textbf{0.0350} & 0.0919 & \underline{0.0467} & \underline{0.1451} & \underline{0.0601} & 0.2360 & \underline{0.0782} \\
\hline
& Ground Truth & 0.0545 & 0.0349 & 0.0912 & 0.0467 & 0.1446 & 0.0603 & 0.2341 & 0.0782 \\
\midrule
\multirow{7}{*}{Beauty} & Gemma-3-4B & \textbf{0.0206} & 0.0110 & 0.0295 & 0.0139 & \textbf{0.0523} & 0.0198 & \textbf{0.0929} & \underline{0.0279} \\
& Gemma-3-12B & 0.0164 & 0.0094 & \underline{0.0318} & 0.0144 & 0.0476 & 0.0186 & 0.0825 & 0.0254 \\
& Gemma-3-27B & 0.0178 & 0.0097 & 0.0318 & 0.0144 & 0.0476 & 0.0185 & 0.0813 & 0.0252 \\
& Qwen2.5-VL-7B & 0.0197 & \underline{0.0112} & 0.0304 & \underline{0.0147} & \underline{0.0499} & \underline{0.0198} & \underline{0.0904} & \textbf{0.0280} \\
& Qwen2.5-VL-32B & 0.0183 & 0.0104 & 0.0304 & 0.0144 & 0.0457 & 0.0186 & 0.0825 & 0.0259 \\
& Qwen2.5-VL-72B & \underline{0.0201} & \textbf{0.0121} & \textbf{0.0337} & \textbf{0.0165} & 0.0467 & \textbf{0.0200} & 0.0789 & 0.0265 \\
\hline
& Ground Truth & 0.0187 & 0.0106 & 0.0314 & 0.0147 & 0.0487 & 0.0192 & 0.0844 & 0.0264 \\
\midrule
\multirow{7}{*}{Electronics} & Gemma-3-4B & \textbf{0.0321} & \textbf{0.0195} & 0.0539 & 0.0267 & \textbf{0.0911} & \textbf{0.0364} & 0.1545 & 0.0493 \\
& Gemma-3-12B & 0.0302 & 0.0186 & 0.0544 & 0.0265 & 0.0893 & 0.0357 & \textbf{0.1553} & 0.0492 \\
& Gemma-3-27B & 0.0315 & 0.0192 & 0.0528 & 0.0263 & 0.0880 & 0.0355 & 0.1523 & 0.0487 \\
& Qwen2.5-VL-7B & 0.0314 & 0.0192 & 0.0544 & 0.0268 & 0.0884 & 0.0357 & 0.1538 & 0.0491 \\
& Qwen2.5-VL-32B & 0.0315 & \underline{0.0194} & \textbf{0.0547} & \textbf{0.0270} & \underline{0.0904} & \underline{0.0364} & 0.1535 & \underline{0.0493} \\
& Qwen2.5-VL-72B & \underline{0.0315} & 0.0193 & \underline{0.0545} & \underline{0.0269} & 0.0897 & 0.0361 & \underline{0.1547} & \textbf{0.0494} \\
\hline
& Ground Truth & 0.0309 & 0.0191 & 0.0541 & 0.0267 & 0.0885 & 0.0358 & 0.1543 & 0.0492 \\
\midrule
\multirow{7}{*}{Home} & Gemma-3-4B & 0.0307 & 0.0215 & 0.0486 & 0.0273 & \underline{0.0843} & \textbf{0.0363} & 0.1540 & \textbf{0.0502} \\
& Gemma-3-12B & \textbf{0.0327} & \textbf{0.0225} & \textbf{0.0505} & \textbf{0.0282} & 0.0806 & 0.0358 & 0.1518 & 0.0501 \\
& Gemma-3-27B & \underline{0.0319} & \underline{0.0217} & \underline{0.0503} & \underline{0.0277} & 0.0825 & 0.0358 & 0.1526 & 0.0498 \\
& Qwen2.5-VL-7B & 0.0303 & 0.0210 & 0.0480 & 0.0268 & 0.0835 & 0.0358 & \underline{0.1540} & 0.0499 \\
& Qwen2.5-VL-32B & 0.0309 & 0.0214 & 0.0476 & 0.0269 & 0.0832 & 0.0359 & 0.1537 & 0.0500 \\
& Qwen2.5-VL-72B & 0.0307 & 0.0213 & 0.0474 & 0.0267 & \textbf{0.0844} & \underline{0.0361} & \textbf{0.1546} & \underline{0.0501} \\
\hline
& Ground Truth & 0.0312 & 0.0215 & 0.0500 & 0.0276 & 0.0803 & 0.0353 & 0.1544 & 0.0501 \\
\midrule
\multirow{7}{*}{Instruments} & Gemma-3-4B & 0.0651 & 0.0423 & 0.1041 & 0.0548 & 0.1540 & 0.0675 & 0.2535 & 0.0873 \\
& Gemma-3-12B & 0.0650 & 0.0423 & \textbf{0.1052} & \textbf{0.0553} & 0.1544 & \underline{0.0677} & 0.2521 & 0.0872 \\
& Gemma-3-27B & \textbf{0.0659} & 0.0422 & \underline{0.1049} & 0.0548 & \underline{0.1552} & 0.0675 & \textbf{0.2551} & \textbf{0.0874} \\
& Qwen2.5-VL-7B & 0.0645 & 0.0419 & 0.1041 & 0.0548 & \textbf{0.1552} & \textbf{0.0677} & 0.2528 & 0.0871 \\
& Qwen2.5-VL-32B & \underline{0.0655} & \textbf{0.0424} & 0.1047 & \underline{0.0550} & 0.1541 & 0.0675 & \underline{0.2537} & \underline{0.0874} \\
& Qwen2.5-VL-72B & 0.0653 & \underline{0.0423} & 0.1028 & 0.0544 & 0.1531 & 0.0672 & 0.2533 & 0.0872 \\
\hline
& Ground Truth & 0.0651 & 0.0425 & 0.1031 & 0.0548 & 0.1539 & 0.0676 & 0.2544 & 0.0877 \\
\midrule
\multirow{7}{*}{Office} & Gemma-3-4B & 0.0525 & 0.0335 & 0.0844 & 0.0438 & 0.1265 & 0.0545 & 0.2072 & 0.0704 \\
& Gemma-3-12B & \underline{0.0535} & \underline{0.0342} & 0.0859 & 0.0447 & 0.1269 & 0.0551 & 0.2070 & \underline{0.0709} \\
& Gemma-3-27B & 0.0533 & 0.0341 & \textbf{0.0862} & 0.0447 & \underline{0.1278} & 0.0552 & \underline{0.2073} & 0.0709 \\
& Qwen2.5-VL-7B & 0.0533 & 0.0340 & 0.0859 & 0.0445 & 0.1269 & 0.0549 & 0.2064 & 0.0706 \\
& Qwen2.5-VL-32B & \textbf{0.0538} & \textbf{0.0344} & 0.0859 & \underline{0.0447} & 0.1273 & \underline{0.0552} & 0.2061 & 0.0708 \\
& Qwen2.5-VL-72B & 0.0529 & 0.0340 & \underline{0.0860} & \textbf{0.0447} & \textbf{0.1281} & \textbf{0.0553} & \textbf{0.2083} & \textbf{0.0711} \\
\hline
& Ground Truth & 0.0533 & 0.0342 & 0.0859 & 0.0448 & 0.1279 & 0.0554 & 0.2065 & 0.0709 \\
\midrule
\multirow{7}{*}{Scientific} & Gemma-3-4B & \textbf{0.0532} & \textbf{0.0345} & 0.0781 & 0.0426 & 0.1167 & 0.0523 & \textbf{0.1913} & \textbf{0.0671} \\
& Gemma-3-12B & 0.0518 & 0.0344 & 0.0771 & 0.0425 & \textbf{0.1186} & \textbf{0.0530} & 0.1887 & \underline{0.0669} \\
& Gemma-3-27B & 0.0493 & 0.0327 & \textbf{0.0814} & \textbf{0.0430} & \underline{0.1184} & 0.0522 & 0.1898 & 0.0664 \\
& Qwen2.5-VL-7B & 0.0519 & 0.0336 & \underline{0.0804} & 0.0428 & 0.1178 & 0.0522 & 0.1905 & 0.0666 \\
& Qwen2.5-VL-32B & \underline{0.0521} & \underline{0.0344} & 0.0786 & \underline{0.0429} & 0.1178 & \underline{0.0528} & 0.1862 & 0.0664 \\
& Qwen2.5-VL-72B & 0.0498 & 0.0329 & 0.0795 & 0.0426 & 0.1182 & 0.0523 & \underline{0.1911} & 0.0667 \\
\hline
& Ground Truth & 0.0541 & 0.0348 & 0.0788 & 0.0428 & 0.1180 & 0.0527 & 0.1908 & 0.0672 \\
\midrule
\multirow{7}{*}{Toys \& Games} & Gemma-3-4B & 0.0373 & 0.0252 & 0.0556 & 0.0311 & \textbf{0.0854} & \underline{0.0386} & \textbf{0.1463} & \textbf{0.0506} \\
& Gemma-3-12B & \underline{0.0378} & 0.0252 & \underline{0.0559} & 0.0311 & 0.0847 & 0.0384 & 0.1426 & 0.0499 \\
& Gemma-3-27B & 0.0371 & 0.0251 & \textbf{0.0571} & \textbf{0.0315} & 0.0851 & 0.0385 & 0.1451 & 0.0504 \\
& Qwen2.5-VL-7B & 0.0376 & \underline{0.0253} & 0.0557 & \underline{0.0312} & \underline{0.0851} & \textbf{0.0386} & 0.1451 & 0.0504 \\
& Qwen2.5-VL-32B & \textbf{0.0382} & \textbf{0.0255} & 0.0553 & 0.0310 & 0.0843 & 0.0383 & 0.1448 & 0.0503 \\
& Qwen2.5-VL-72B & 0.0373 & 0.0253 & 0.0558 & 0.0312 & 0.0848 & 0.0385 & \underline{0.1453} & \underline{0.0505} \\
\hline
& Ground Truth & 0.0376 & 0.0255 & 0.0559 & 0.0314 & 0.0851 & 0.0388 & 0.1452 & 0.0506 \\
\midrule
\multirow{7}{*}{Video Games} & Gemma-3-4B & 0.0621 & \underline{0.0406} & 0.0990 & 0.0525 & \textbf{0.1541} & \textbf{0.0666} & 0.2519 & \underline{0.0861} \\
& Gemma-3-12B & 0.0605 & 0.0392 & 0.0968 & 0.0510 & 0.1514 & 0.0648 & 0.2497 & 0.0844 \\
& Gemma-3-27B & \underline{0.0623} & 0.0403 & 0.0991 & 0.0522 & 0.1521 & 0.0657 & \underline{0.2532} & 0.0859 \\
& Qwen2.5-VL-7B & 0.0608 & 0.0394 & 0.0966 & 0.0510 & 0.1507 & 0.0647 & 0.2507 & 0.0847 \\
& Qwen2.5-VL-32B & 0.0620 & 0.0404 & \textbf{0.1001} & \textbf{0.0528} & \underline{0.1539} & \underline{0.0664} & \textbf{0.2535} & \textbf{0.0863} \\
& Qwen2.5-VL-72B & \textbf{0.0630} & \textbf{0.0408} & \underline{0.0999} & \underline{0.0527} & 0.1534 & 0.0663 & 0.2518 & 0.0859 \\
\hline
& Ground Truth & 0.0623 & 0.0404 & 0.0985 & 0.0521 & 0.1523 & 0.0658 & 0.2508 & 0.0855 \\
\bottomrule
\end{tabular}

\label{tab:performance_freedom_t_to_i}
\end{table*}
\clearpage




\bibliographystyle{elsarticle-num-names}
\bibliography{cas-refs}

@inproceedings{fu2024exploring,
  title={Exploring adapter-based transfer learning for recommender systems: Empirical studies and practical insights},
  author={Fu, Junchen and Yuan, Fajie and Song, Yu and Yuan, Zheng and Cheng, Mingyue and Cheng, Shenghui and Zhang, Jiaqi and Wang, Jie and Pan, Yunzhu},
  booktitle={Proceedings of the 17th ACM international conference on web search and data mining},
  pages={208--217},
  year={2024}
}

@article{fu2026llmpopcorn,
  title={LLMpopcorn: exploring LLMs as assistants for popular micro-video generation},
  author={Fu, Junchen and Ge, Xuri and Zheng, Kaiwen and Karatzoglou, Alexandros and Arapakis, Ioannis and Xin, Xin and Ni, Yongxin and Jose, Joemon M},
journal={arXiv preprint arXiv:2502.12945
        
        
        
        
        
        
        
        },
  year={2026}
}

@article{fu2025crossan,
  title={CROSSAN: Towards Efficient and Effective Adaptation of Multiple Multimodal Foundation Models for Sequential Recommendation},
  author={Fu, Junchen and Ni, Yongxin and Jose, Joemon M and Arapakis, Ioannis and Zheng, Kaiwen and Li, Youhua and Ge, Xuri},
  journal={arXiv preprint arXiv:2504.10307
        
        
        
        
        
        
        
        
        
        
        
        },
  year={2025}
}

@article{fu2025efficient,
  title={Efficient and effective adaptation of multimodal foundation models in sequential recommendation},
  author={Fu, Junchen and Ge, Xuri and Xin, Xin and Karatzoglou, Alexandros and Arapakis, Ioannis and Zheng, Kaiwen and Ni, Yongxin and Joemon, Joemon M Jose},
  journal={IEEE Transactions on Knowledge and Data Engineering},
  year={2025},
  publisher={IEEE}
}

@inproceedings{fu2024iisan,
  title={IISAN: Efficiently adapting multimodal representation for sequential recommendation with decoupled PEFT},
  author={Fu, Junchen and Ge, Xuri and Xin, Xin and Karatzoglou, Alexandros and Arapakis, Ioannis and Wang, Jie and Jose, Joemon M},
  booktitle={Proceedings of the 47th International ACM SIGIR Conference on Research and Development in Information Retrieval},
  pages={687--697},
  year={2024}
}

@article{chen2025adaptive,
  title={Adaptive Latent Disease State Learning for Multimodal Alzheimer’s Disease Biomarker Detection with Missing Modalities},
  author={Chen, Zhi and Zhang, Fengli and Zhang, Yun and Zhu, Jiajing and Li, Qiaoqin and Liu, Yongguo},
  journal={Pattern Recognition},
  pages={112389},
  year={2025},
  publisher={Elsevier}
}

@article{qiu2025mimar,
  title={MIMAR-OSA: Enhancing obstructive sleep apnea diagnosis through multimodal data integration and missing modality reconstruction},
  author={Qiu, Xihe and Wei, Yingchen and Tan, Xiaoyu and Xu, Weidi and Wang, Haodong and Ma, Jingru and Huang, Jingjing and Fang, Zhijun},
  journal={Pattern Recognition},
  pages={111917},
  year={2025},
  publisher={Elsevier}
}

@article{zhao2025mce,
  title={MCE: Towards a General Framework for Handling Missing Modalities under Imbalanced Missing Rates},
  author={Zhao, Binyu and Zhang, Wei and Zou, Zhaonian},
  journal={Pattern Recognition},
  pages={112591},
  year={2025},
  publisher={Elsevier}
}

@inproceedings{zhu2021knowledge,
  title={Knowledge perceived multi-modal pretraining in e-commerce},
  author={Zhu, Yushan and Zhao, Huaixiao and Zhang, Wen and Ye, Ganqiang and Chen, Hui and Zhang, Ningyu and Chen, Huajun},
  booktitle={Proceedings of the 29th ACM international conference on multimedia},
  pages={2744--2752},
  year={2021}
}

@article{wang2004image,
  title={Image quality assessment: from error visibility to structural similarity},
  author={Wang, Zhou and Bovik, Alan C and Sheikh, Hamid R and Simoncelli, Eero P},
  journal={IEEE transactions on image processing},
  volume={13},
  number={4},
  pages={600--612},
  year={2004},
  publisher={IEEE}
}

@inproceedings{zhang2018unreasonable,
  title={The unreasonable effectiveness of deep features as a perceptual metric},
  author={Zhang, Richard and Isola, Phillip and Efros, Alexei A and Shechtman, Eli and Wang, Oliver},
  booktitle={Proceedings of the IEEE conference on computer vision and pattern recognition},
  pages={586--595},
  year={2018}
}

@article{wang2020measurement,
  title={Measurement of text similarity: a survey},
  author={Wang, Jiapeng and Dong, Yihong},
  journal={Information},
  volume={11},
  number={9},
  pages={421},
  year={2020},
  publisher={MDPI}
}

@article{tan2013perceptually,
  title={A perceptually relevant MSE-based image quality metric},
  author={Tan, Hui Li and Li, Zhengguo and Tan, Yih Han and Rahardja, Susanto and Yeo, Chuohuo},
  journal={IEEE Transactions on Image Processing},
  volume={22},
  number={11},
  pages={4447--4459},
  year={2013},
  publisher={IEEE}
}

@article{tata2007estimating,
  title={Estimating the selectivity of tf-idf based cosine similarity predicates},
  author={Tata, Sandeep and Patel, Jignesh M},
  journal={ACM Sigmod Record},
  volume={36},
  number={2},
  pages={7--12},
  year={2007},
  publisher={ACM New York, NY, USA}
}

@article{shen2020multi,
  title={Multi-domain image completion for random missing input data},
  author={Shen, Liyue and Zhu, Wentao and Wang, Xiaosong and Xing, Lei and Pauly, John M and Turkbey, Baris and Harmon, Stephanie Anne and Sanford, Thomas Hogue and Mehralivand, Sherif and Choyke, Peter L and others},
  journal={IEEE transactions on medical imaging},
  volume={40},
  number={4},
  pages={1113--1122},
  year={2020},
  publisher={IEEE}
}

@article{elmore2001euclidean,
  title={Euclidean distance as a similarity metric for principal component analysis},
  author={Elmore, Kimberly L and Richman, Michael B},
  journal={Monthly weather review},
  volume={129},
  number={3},
  pages={540--549},
  year={2001}
}

@inproceedings{ghalandari2022efficient,
  title={Efficient Unsupervised Sentence Compression by Fine-tuning Transformers with Reinforcement Learning},
  author={Ghalandari, Demian and Hokamp, Chris and Ifrim, Georgiana},
  booktitle={Proceedings of the 60th Annual Meeting of the Association for Computational Linguistics (Volume 1: Long Papers)},
  pages={1267--1280},
  year={2022}
}

@inproceedings{vaze2023genecis,
  title={Genecis: A benchmark for general conditional image similarity},
  author={Vaze, Sagar and Carion, Nicolas and Misra, Ishan},
  booktitle={Proceedings of the IEEE/CVF Conference on Computer Vision and Pattern Recognition},
  pages={6862--6872},
  year={2023}
}

@inproceedings{hore2010image,
  title={Image quality metrics: PSNR vs. SSIM},
  author={Hore, Alain and Ziou, Djemel},
  booktitle={2010 20th international conference on pattern recognition},
  pages={2366--2369},
  year={2010},
  organization={IEEE}
}

@inproceedings{wang2020transmodality,
  title={Transmodality: An end2end fusion method with transformer for multimodal sentiment analysis},
  author={Wang, Zilong and Wan, Zhaohong and Wan, Xiaojun},
  booktitle={Proceedings of the web conference 2020},
  pages={2514--2520},
  year={2020}
}

@article{shao2024deepseekmath,
  title={Deepseekmath: Pushing the limits of mathematical reasoning in open language models},
  author={Shao, Zhihong and Wang, Peiyi and Zhu, Qihao and Xu, Runxin and Song, Junxiao and Bi, Xiao and Zhang, Haowei and Zhang, Mingchuan and Li, YK and Wu, Yang and others},
  journal={arXiv preprint arXiv:2402.03300
        
        
        
        
        
        
        
        
        
        
        
        },
  year={2024}
}

@article{lan2025survey,
  title={A Survey of Automatic Evaluation Methods on Text, Visual and Speech Generations},
  author={Lan, Tian and Zhou, Yang-Hao and Ma, Zi-Ao and Sun, Fanshu and Sun, Rui-Qing and Luo, Junyu and Tu, Rong-Cheng and Huang, Heyan and Xu, Chen and Wu, Zhijing and others},
  journal={arXiv preprint arXiv:2506.10019
        
        
        
        
        
        },
  year={2025}
}

@article{zhang2019bertscore,
  title={Bertscore: Evaluating text generation with bert},
  author={Zhang, Tianyi and Kishore, Varsha and Wu, Felix and Weinberger, Kilian Q and Artzi, Yoav},
  journal={arXiv preprint arXiv:1904.09675
        
        
        
        },
  year={2019}
}

@inproceedings{zhou2023tale,
  title={A tale of two graphs: Freezing and denoising graph structures for multimodal recommendation},
  author={Zhou, Xin and Shen, Zhiqi},
  booktitle={Proceedings of the 31st ACM international conference on multimedia},
  pages={935--943},
  year={2023}
}

@inproceedings{zhou2023bootstrap,
  title={Bootstrap latent representations for multi-modal recommendation},
  author={Zhou, Xin and Zhou, Hongyu and Liu, Yong and Zeng, Zhiwei and Miao, Chunyan and Wang, Pengwei and You, Yuan and Jiang, Feijun},
  booktitle={Proceedings of the ACM web conference 2023},
  pages={845--854},
  year={2023}
}

@inproceedings{he2016vbpr,
  title={VBPR: visual bayesian personalized ranking from implicit feedback},
  author={He, Ruining and McAuley, Julian},
  booktitle={Proceedings of the AAAI conference on artificial intelligence},
  volume={30},
  number={1},
  year={2016}
}

@inproceedings{hou2022towards,
  title={Towards universal sequence representation learning for recommender systems},
  author={Hou, Yupeng and Mu, Shanlei and Zhao, Wayne Xin and Li, Yaliang and Ding, Bolin and Wen, Ji-Rong},
  booktitle={Proceedings of the 28th ACM SIGKDD conference on knowledge discovery and data mining},
  pages={585--593},
  year={2022}
}

@inproceedings{zhang2025hierarchical,
  title={Hierarchical Time-Aware Mixture of Experts for Multi-Modal Sequential Recommendation},
  author={Zhang, Shengzhe and Chen, Liyi and Shen, Dazhong and Wang, Chao and Xiong, Hui},
  booktitle={Proceedings of the ACM on Web Conference 2025},
  pages={3672--3682},
  year={2025}
}

@inproceedings{zhou2023mmrec,
  title={Mmrec: Simplifying multimodal recommendation},
  author={Zhou, Xin},
  booktitle={Proceedings of the 5th ACM International Conference on Multimedia in Asia Workshops},
  pages={1--2},
  year={2023}
}

@inproceedings{ke2025knowledge,
  title={Knowledge Bridger: Towards Training-Free Missing Modality Completion},
  author={Ke, Guanzhou and He, Shengfeng and Wang, Xiaoli and Wang, Bo and Chao, Guoqing and Zhang, Yuanyang and Xie, Yi and Su, Hexing},
  booktitle={Proceedings of the Computer Vision and Pattern Recognition Conference},
  pages={25864--25873},
  year={2025}
}

@inproceedings{wang2018lrmm,
  title={LRMM: Learning to Recommend with Missing Modalities},
  author={Wang, Cheng and Niepert, Mathias and Li, Hui},
  booktitle={Proceedings of the 2018 Conference on Empirical Methods in Natural Language Processing},
  pages={3360--3370},
  year={2018}
}

@inproceedings{wang2023mpkgac,
  title     = {MPKGAC: Multimodal Product Attribute Completion in E-commerce},
  author    = {Wang, Kai and Shao, Jianzhi and Zhang, Tao and Chen, Qijin and Huo, Chengfu},
  booktitle = {Companion Proceedings of the ACM Web Conference 2023},
  pages     = {336--340},
  year      = {2023},
  doi       = {10.1145/3543873.3584623}
}

@article{vanbuuren2011mice,
  title   = {mice: Multivariate imputation by chained equations in R},
  author  = {van Buuren, Stef and Groothuis-Oudshoorn, Karin},
  journal = {Journal of Statistical Software},
  volume  = {45},
  number  = {3},
  pages   = {1--67},
  year    = {2011},
  doi     = {10.18637/jss.v045.i03}
}

@article{stekhoven2012missforest,
  title   = {MissForest---non-parametric missing value imputation for mixed-type data},
  author  = {Stekhoven, Daniel J. and B{\"u}hlmann, Peter},
  journal = {Bioinformatics},
  volume  = {28},
  number  = {1},
  pages   = {112--118},
  year    = {2012},
  doi     = {10.1093/bioinformatics/btr597}
}

@inproceedings{radford2021clip,
  title     = {Learning Transferable Visual Models From Natural Language Supervision},
  author    = {Radford, Alec and Kim, Jong Wook and Hallacy, Chris and Ramesh, Aditya and Goh, Gabriel and Agarwal, Sandhini and Sastry, Girish and Askell, Amanda and Mishkin, Pamela and Clark, Jack and Krueger, Gretchen and Sutskever, Ilya},
  booktitle = {Proceedings of the 38th International Conference on Machine Learning},
  pages     = {8748--8763},
  year      = {2021},
  publisher = {PMLR}
}

@inproceedings{li2020aspect,
  title={Aspect-aware multimodal summarization for chinese e-commerce products},
  author={Li, Haoran and Yuan, Peng and Xu, Song and Wu, Youzheng and He, Xiaodong and Zhou, Bowen},
  booktitle={Proceedings of the AAAI conference on artificial intelligence},
  volume={34},
  number={05},
  pages={8188--8195},
  year={2020}
}

@inproceedings{malitesta2024we,
  title={Do We Really Need to Drop Items with Missing Modalities in Multimodal Recommendation?},
  author={Malitesta, Daniele and Rossi, Emanuele and Pomo, Claudio and Di Noia, Tommaso and Malliaros, Fragkiskos D},
  booktitle={Proceedings of the 33rd ACM International Conference on Information and Knowledge Management},
  pages={3943--3948},
  year={2024}
}

@inproceedings{ma2021smil,
  title     = {{SMIL}: Multimodal Learning with Severely Missing Modality},
  author    = {Ma, Mengmeng and Ren, Jian and Zhao, Long and Tulyakov, Sergey and Wu, Cathy and Peng, Xi},
  booktitle = {Proceedings of the 35th AAAI Conference on Artificial Intelligence},
  year      = {2021},
  pages     = {2302--2310}
}

@inproceedings{ma2022robust,
  title     = {Are Multimodal Transformers Robust to Missing Modality?},
  author    = {Ma, Mengmeng and Ren, Jian and Zhao, Long and Testuggine, Davide and Peng, Xi},
  booktitle = {CVPR},
  year      = {2022},
  pages     = {18156--18165}
}

@inproceedings{lee2023prompting,
  title     = {Multimodal Prompting with Missing Modalities for Visual Recognition},
  author    = {Lee, Yi-Lun and Tsai, Yi-Hsuan and Chiu, Wei-Chen and Lee, Chen-Yu},
  booktitle = {CVPR},
  year      = {2023},
  pages     = {14943--14952}
}

@inproceedings{shi2019adaptive,
  title     = {Adaptive Feature Sampling for Recommendation with Missing Content Feature Values},
  author    = {Shi, Shaoyun and Zhang, Min and Yu, Xinxing and Zhang, Yongfeng and Hao, Bin and Liu, Yiqun and Ma, Shaoping},
  booktitle = {CIKM},
  year      = {2019},
  pages     = {1451--1460}
}

@article{liu2022attribute,
  title   = {An Attribute-Aware Attentive Graph Convolutional Network for Attribute Missing in Recommendation},
  author  = {Liu, Fan and Cheng, Zhiyong and Zhu, Lei and Liu, Chenghao and Nie, Liqiang},
  journal = {IEEE Transactions on Knowledge and Data Engineering},
  volume  = {34},
  number  = {9},
  year    = {2022},
  pages   = {4077--4088}
}

@inproceedings{kim2025dgmrec,
  title     = {Disentangling and Generating Modalities for Recommendation in Missing Modality Scenarios},
  author    = {Kim, Jiwan and Kang, Hongseok and Kim, Sein and Kim, Kibum and Park, Chanyoung},
  booktitle = {SIGIR},
  year      = {2025},
  note      = {In press}
}

\end{document}